 \documentclass[pra,amsmath,amssymb,twocolumn, showpacs, superscriptaddress,10pt]{revtex4-1}

\usepackage{amsmath}
\usepackage{hyperref}
\usepackage{graphicx}
\usepackage{amsfonts}
\usepackage{amsthm}
\usepackage{cases}
\usepackage{bm}

\usepackage[normalem]{ulem}

\usepackage{color}
\definecolor{Blue}{rgb}{0.00, 0.00, 1.00}
\definecolor{Red}{rgb}{1.00, 0.00, 0.00}
\definecolor{Green}{rgb}{0.00, 0.60, 0.00}

\hypersetup{
    colorlinks=true,       
    linkcolor=red,          
    citecolor=blue,        
    filecolor=magenta,      
    urlcolor=cyan           
}

\newcommand{\nn}{\nonumber}
\newcommand{\be}{\begin{equation}}
\newcommand{\ee}{\end{equation}}
\newcommand{\bea}{\begin{eqnarray}}
\newcommand{\eea}{\end{eqnarray}}

\newcommand{\beq}{\begin{equation}}
\newcommand{\eeq}{\end{equation}}
\newcommand{\beqn}{\begin{eqnarray}}
\newcommand{\eeqn}{\end{eqnarray}}

\DeclareMathOperator{\Tr}{Tr}

\begin{document}

\title{Counting statistics for non-interacting fermions in a rotating trap }

\author{Naftali R. Smith}
\email{naftalismith@gmail.com}
\affiliation{Laboratoire de Physique de l’Ecole Normale Sup\'erieure, CNRS, ENS \& Universit\'e PSL, Sorbonne Universit\'e,
Universit\'e de Paris, 75005 Paris, France}
\affiliation{Department of Solar Energy and Environmental Physics, Blaustein Institutes for Desert Research, Ben-Gurion University of the Negev, Sede Boqer Campus, 8499000, Israel}
\author{Pierre Le Doussal}
\email{pierre.ledoussal@phys.ens.fr}
\affiliation{Laboratoire de Physique de l’Ecole Normale Sup\'erieure, CNRS, ENS \& Universit\'e PSL, Sorbonne Universit\'e,
Universit\'e de Paris, 75005 Paris, France}
\author{Satya N. \surname{Majumdar}}
\email{satya.majumdar@universite-paris-saclay.fr}
\affiliation{Universit{\'e} Paris-Saclay, CNRS, LPTMS, 91405, Orsay, France}
\author{Gr\'egory \surname{Schehr}}
\email{schehr@lpthe.jussieu.fr}
\affiliation{Sorbonne Universit{\'e}, Laboratoire de Physique Th{\'e}orique et Hautes Energies, CNRS UMR 7589, 4 Place Jussieu, 75252 Paris Cedex 05, France}
\date{\today}

\begin{abstract}
We study the ground state of $N \gg 1$ noninteracting fermions in a two-dimensional harmonic trap rotating at angular frequency $\Omega>0$. 
The support of the density of the Fermi gas is a disk of radius $R_e$.
We calculate the variance of the number of fermions ${\cal N}_R$ inside a disk of radius $R$ centered at the origin for $R$ in the bulk of the Fermi gas. We find rich and interesting behaviours in two different scaling regimes: (i) $\Omega / \omega <1 $ and (ii) $1 - \Omega / \omega = O(1/N)$, where $\omega$ is the angular frequency of the oscillator. In the first regime (i) we find that ${\rm Var}\,{\cal N}_{R}\simeq\left(A\log N+B\right)\sqrt{N}$ and we calculate $A$ and $B$ as functions of $R/R_e$, $\Omega$ and $\omega$. We also predict the higher cumulants of ${\cal N}_{R}$ and the bipartite entanglement entropy of the disk with the rest of the system.
In the second regime (ii), the mean fermion density exhibits a staircase form, with discrete plateaus corresponding to filling $k$ successive Landau levels, as found in previous studies. Here, we show that ${\rm Var}\,{\cal N}_{R}$ is a discontinuous piecewise linear function of $\sim (R/R_e) \sqrt{N}$ within each plateau, with coefficients that we calculate exactly, and with steps whose precise shape we obtain for any $k$. 
We argue that a similar piecewise linear behavior extends to all the cumulants of ${\cal N}_{R}$ and to the entanglement entropy. 
We show that these results match smoothly at large $k$ with the above results
for $\Omega/\omega=O(1)$. These findings are nicely confirmed by numerical simulations.
Finally, we uncover a universal behavior of ${\rm Var}\,{\cal N}_{R}$ near the fermionic edge. 
We extend our results to a three-dimensional geometry,
where an additional confining potential is applied in the $z$ direction.

\end{abstract}



\maketitle

\section{Introduction}

Noninteracting fermions confined by a trapping potential is a topic which has attracted much interest over the recent years.
This study is motivated by recent progress on the experimental side, where systems of cold atoms have been realized, manipulated and measured with quantum gas microscopes at high resolution \cite{Fermicro1,Fermicro2,Fermicro3,Pauli, FermicroYang21, BDZ08,flattrap}. In particular, quantum gas microscopy enables one to observe many-body systems at the single-atom precision and in principle allows to measure the counting statistics of cold gases \cite{FCS_malossi,FCS_Schempo,omran}.
The noninteracting limit is experimentally relevant since the interactions between the particles can be tuned in experiments.
Studies on the theoretical side have focused on the number density, its correlations and related properties, in real space and in momentum space \cite{V12, Eisler1, MMSV14, DeanEPL2015, MMSV16,  DeanPLDReview, RMSZG17, DeanReview2019,Deleporte21}. Systems of noninteracting fermions display rich and interesting behaviour and nontrivial fluctuations even at zero temperature due to the Pauli exclusion principle.
The Local Density Approximation (LDA) correctly describes the density fluctuations in the bulk of the Fermi gas \cite{BR1997,Castin}, but it breaks down near the edges \cite{koh98, Eisler1, DeanEPL2015, DeanPLDReview}.
Remarkably, for particular potentials in 1d, there exist exact mappings between the positions of the fermions in the ground state and the eigenvalues of certain random matrix ensembles \cite{Eisler1, MMSV14} -- see \cite{DeanReview2019} for a recent review. Using this connection, the density and its correlations near the edge were calculated and shown to be universal with respect to the trapping potential \cite{Eisler1,DeanPLDReview, DeanReview2019, LaCroix17, Wigner18, Multicritical18}.
However, the connection to random matrix theory (RMT) does not hold generically in $d>1$.

In this paper, we consider $N$ noninteracting spinless fermions in a confining potential which is invariant under a rotation around the $z$-axis. The whole system is then put in rotation around the $z$ axis
with angular frequency $\Omega>0$.
This system was studied experimentally  \cite{ho2000rapidly,schweikhard2004rapidly,aftalion2005vortex}, and later theoretically \cite{fetter2009rotating, cooper2008rapidly,LMG19,KulkarniRotating2020}, {see also the very recent work \cite{DHL22}.}
In the $2d$ geometry the fermions live in the $xy$ plane and feel the external potential
$V(r)$, where ${\bf r}=(x,y)$ and $r=|{\bf r}|$. In the rotating frame at angular frequency $\Omega$ 
the many-body Hamiltonian $\hat {\cal H}_N$ is time-independent and given 
by $\hat {\cal H}_N = \sum_{i=1}^N \hat H_i$ where the single-particle Hamiltonian 
$\hat H_i = \hat H(\hat {\bf p}_i, \hat {\bf r}_i)$ reads (in units where the mass $m=1$
and $\hbar=1$)
\cite{landau1980statistical,leggett2006quantum}
\bea \label{Ham2d} 
 H({\bf p},{\bf r}) &=& \frac{p^2}{2} + V(r) - \Omega L_z \nn\\
& =& \frac{1}{2} ({\bf p} - {\bf A})^2 + V(r) - \frac{\Omega^2}{2} r^2 
\eea
where $p=|{\bf p}|$, $L_z= x p_y - y p_x$ is the $z$-component of the angular momentum
and ${\bf A}=\Omega (y,-x)$. These two expressions are equivalent and the
second shows the Coriolis vector potential ${\bf A}$ and the centrifugal potential
$- \frac{\Omega^2}{2} r^2$ which tends to destabilize the fermion gas.

This problem was studied in \cite{LMG19} for the ground state of $N$ fermions in 
a harmonic trap $V(r)= \frac{1}{2} \omega^2 r^2$.
The focus was on the case $\omega \to \Omega^+$ where the trapping force exactly balances the centrifugal force
and the problem becomes equivalent to fermions in a $2d$-plane and in the presence of a magnetic field perpendicular to that plane. The physics is thus the
one of the lowest Landau level (LLL). In this case, the positions of the fermions can be mapped to the eigenvalue (in the complex $x+i y$ plane) of the Ginibre ensemble of random matrices \cite{Forrester}. In the large $N$ limit, the Fermi gas consequently forms a circular droplet with a uniform density
$\rho = \frac{1}{\pi}$, and a radius $R_e=\sqrt{N}$. In that case, the full counting statistics (FCS) could be obtained exactly. In particular it was found \cite{LMG19}
that all cumulants of the number ${\cal N}_R$ of fermions inside a disk of radius $R$ (centered on the origin $r=0$) grows as $\sim R$ for large $R$. This was recently proved rigorously in \cite{CharlierGin} where higher order asymptotics were obtained.   

This is in marked contrast with the behavior of e.g free fermions in $2d$ (with $\Omega=0$), where the variance of ${\cal N}_R$ grows as
$R \log R$, a result which was extended and refined in the presence of an arbitrary external potential
$V(r)$ in our recent work \cite{UsCounting2020}. Since the LLL is a very special case with infinite degeneracy, one may wonder how the 
variance crosses over from linear in $R$ to the $R \log R$ behavior when one departs from the LLL situation. In \cite{KulkarniRotating2020} the case of a harmonic trap with an additional repulsive potential, $V(r)= \frac{1}{2} \omega^2 r^2 + \frac{\gamma}{2 r^2}$, with $\gamma>0$, was studied in the regime near the LLL where
$\Omega/\omega \to 1$. In that case 
the density is not uniform anymore, but exhibits plateaus at discrete values. It was shown that a rich structure emerges, where
a hole is created near the origin while an additional "wedding cake" discrete layers appear around this hole. However the FCS was not studied there.

\begin{figure*}[ht]
\centering
\includegraphics[angle=0,width=0.49\linewidth]{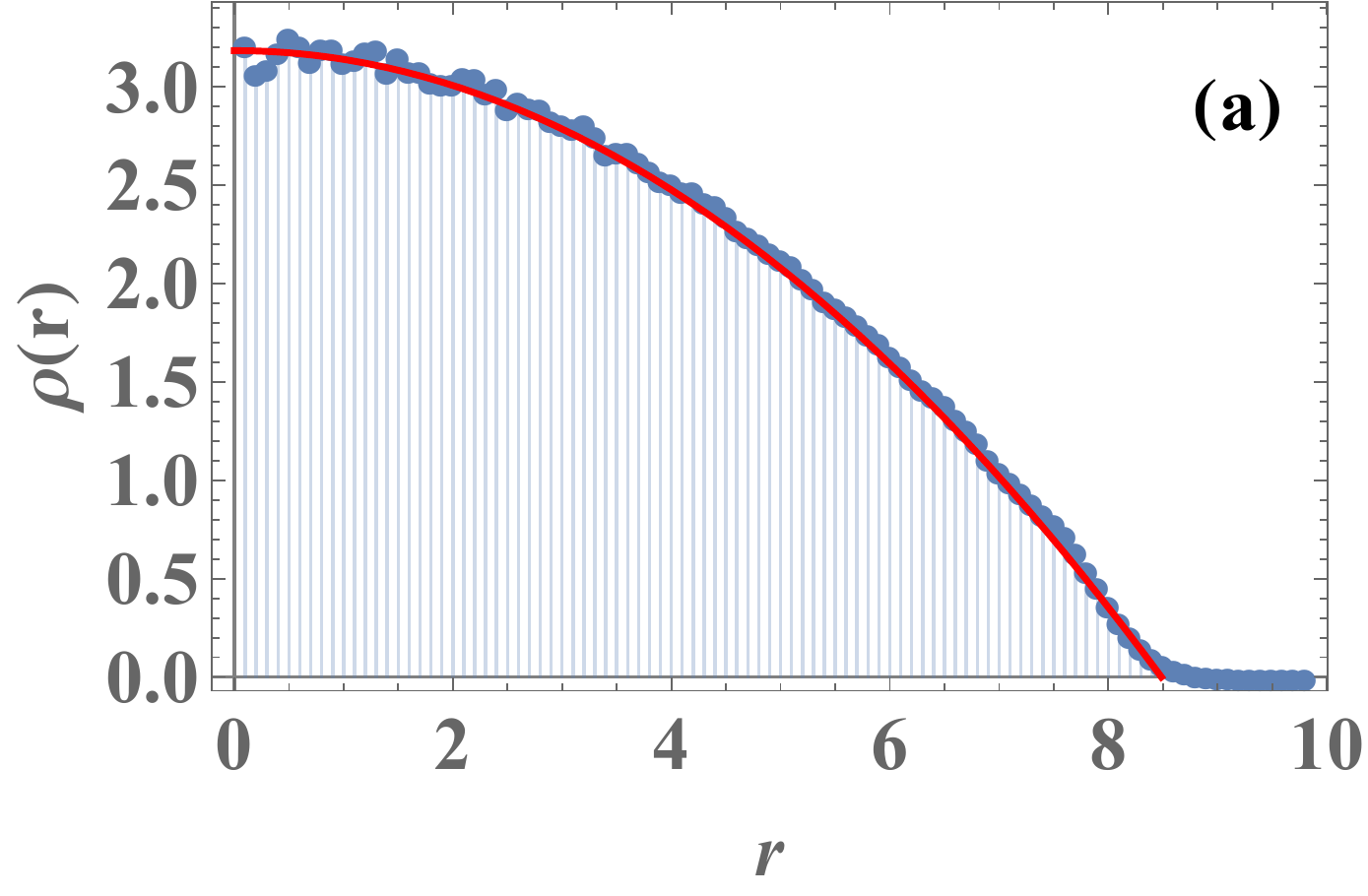}
\includegraphics[angle=0,width=0.49\linewidth]{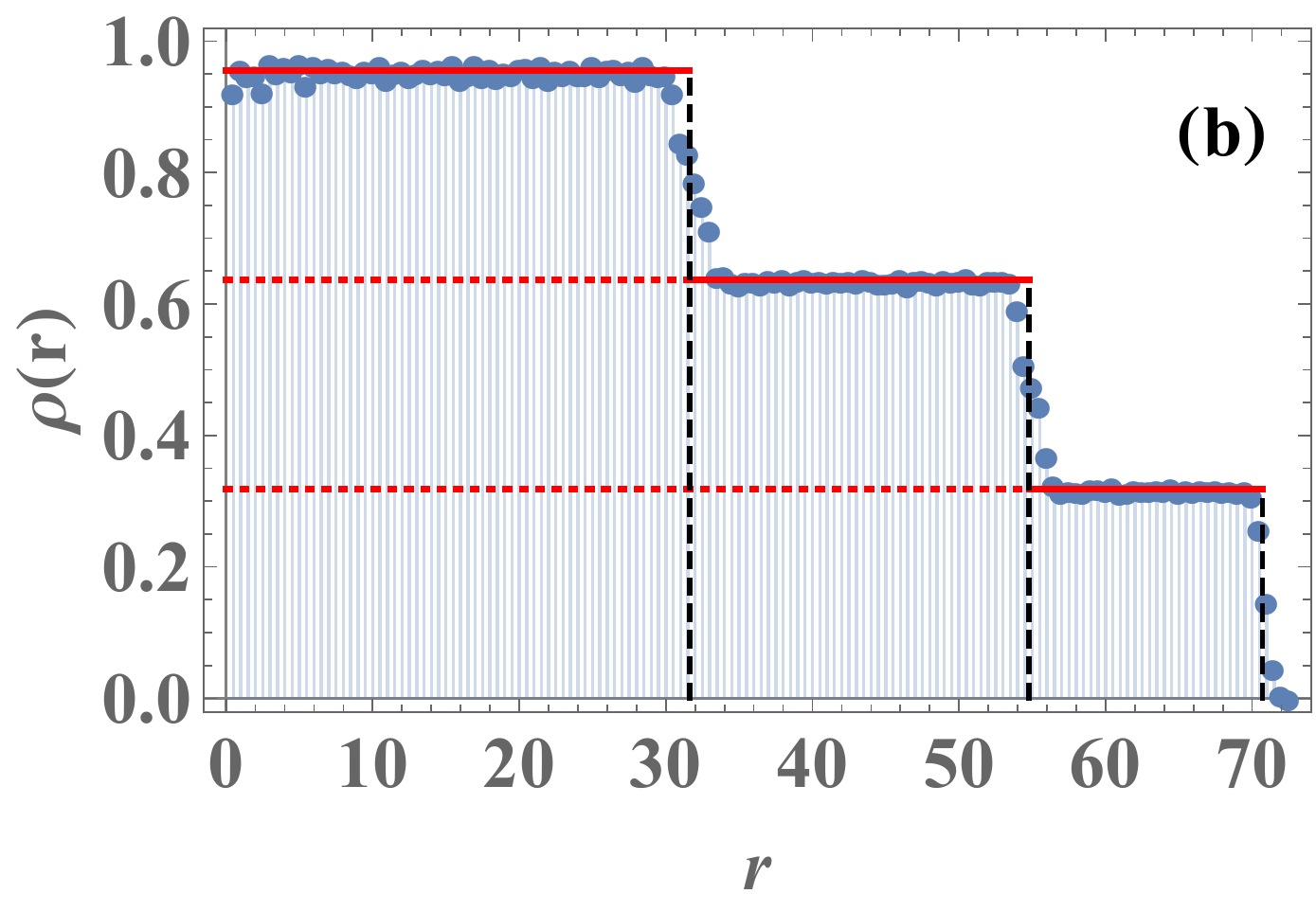}
\caption{Density of the particles for $N$ fermions in a rotating harmonic trap in the two regimes 
(a) $1-\Omega=O(1)$  and (b) $1-\Omega=O(1/N)$.
Solid red line: theoretical predictions, Eq.~\eqref{eq:DensityOmegaO1} in (a) Eq.~\eqref{densfinal} in (b). In (b), the black dashed lines correspond to $R=\sqrt{\ell_k}$, and the red dotted lines are the continuations of the formula \eqref{densfinal} to $r=0$. Blue bars: the density computed empirically from numerical simulations. Parameters are $\Omega = 2/3$, $\mu = 20$ corresponding to a total of $N=366$ fermions in (a), and $\Omega = 0.999$ and Fermi energy $\mu=6$, corresponding to $N=9003$ in (b). The number of samples was $5000$ in (a) and $200$ in (b).}
\label{FigDensity}
\end{figure*}

In this paper we study the FCS of ${\cal N}_R$ for arbitrary values of $\Omega/\omega<1$ (and $\gamma=0$). We will explore the entire
crossover from the regime $1-\frac{\Omega}{\omega} = O(1/N)$ to the regime $\frac{\Omega}{\omega}=O(1)$ and obtain explicit formula for the variance, and 
in some cases for the higher cumulants of ${\cal N}_R$. We start by presenting, in Section \ref{sec:genf}, the general structure of the
ground state of $N$ noninteracting fermions in the rotating harmonic trap, and discuss the various scaling regimes in the large $N$ limit.
To obtain the FCS in the regime $\frac{\Omega}{\omega}=O(1)$ we extend in Section~\ref{sec:3} the method of our previous work at $\Omega=0$ \cite{UsCounting2020} to a nonzero rotating frequency $\Omega > 0$. In this regime the radius $R_e$ of the Fermi gas is given by \eqref{def_Re} and its density by \eqref{eq:DensityOmegaO1}. We obtain the explicit formula for ${\rm Var}\, {\cal N}_R$
at large $N$. It is displayed in Eqs. \eqref{eq:VarNROmegaO1}-\eqref{eq:BROmega} as a function 
of the Fermi energy $\mu$, which is related to $N$ via Eq. \eqref{rel_Nmu} (note that below we set $\omega=1$). 
The higher cumulants of ${\cal N}_R$ are given in \eqref{cumulants}. From these cumulants one obtains the bipartite entanglement entropy of the disk with the rest of the system, given in Eq. \eqref{entropy}.
These results are valid for the bulk of the Fermi gas. We finally obtain the 
universal behavior of ${\rm Var}\,{\cal N}_{R}$ near the fermionic edge, see Eq. \eqref{eq:VarNREdge}.  
In Section \ref{sec:LLL} we study the regime $1- \frac{\Omega}{\omega}=O(1/N)$, i.e., the vicinity of 
the LLL where a finite number of Landau levels are occupied. In this regime the mean fermion density exhibits a staircase form, 
with discrete plateaus at values $\rho = \frac{k}{\pi}$ corresponding to filling $k$ successive Landau levels, as found in 
\cite{LMG19,KulkarniRotating2020}. Here we compute the variance ${\rm Var}\,{\cal N}_{R}$, which is found to be a discontinuous piecewise linear function $\sqrt{2} C_k R$ as given in \eqref{eq:VarNR}, where $R=O(R_e)=O(\sqrt{N})$, and the coefficients $C_k$ associated to the $k$-th Landau level
are given in \eqref{Ck_exp}. The result for $C_1$ agrees with \cite{LMG19} using a different method. 
The discontinuities of ${\rm Var}\,{\cal N}_{R}$ are smeared on the smaller scale $O(1)$ near the steps, and we 
obtain their analytical shape for any $k$ in Eq. \eqref{eq:VarNRStep}. 
We argue that a similar piecewise linear behavior extends to all the cumulants of ${\cal N}_{R}$.
In Section \ref{sec:matching} we show that the results for the variance in the bulk in the regime
$1- \frac{\Omega}{\omega}=O(1/N)$ match smoothly at large $k$ with the above results
for $\Omega/\omega=O(1)$. These analytical results are corroborated by thorough numerical
simulations relying on mappings to random matrix models (see Appendix \ref{app:sim} for details). This allows
to compute numerically the density (see Fig. \ref{FigDensity}) as well as the variance (see Fig. \ref{FigVarNR}) in both regimes. 
Furthermore, in Section \ref{sec:3d} we show how our results can be extended to a three-dimensional geometry,
where an additional confining potential is applied in the $z$ direction
described by the Hamiltonian
\be
\label{eq:H3d}
 H({\bf p},p_z,{\bf r},z) = \frac{p^2}{2} + \frac{p_z^2}{2} + V(r)  + U(z) - \Omega L_z
\ee 
where $U(z)$ is a confining potential in the $z$ direction. 
This extension is important for experimental applications 
where particles are confined to remain in the vicinity of the plane $z=0$.
Finally, further technical details are given in the Appendices \ref{sec:semiclassical} to \ref{sec:higher}.

\begin{figure}[h!t]
\centering
\includegraphics[angle=0,width=0.6\linewidth]{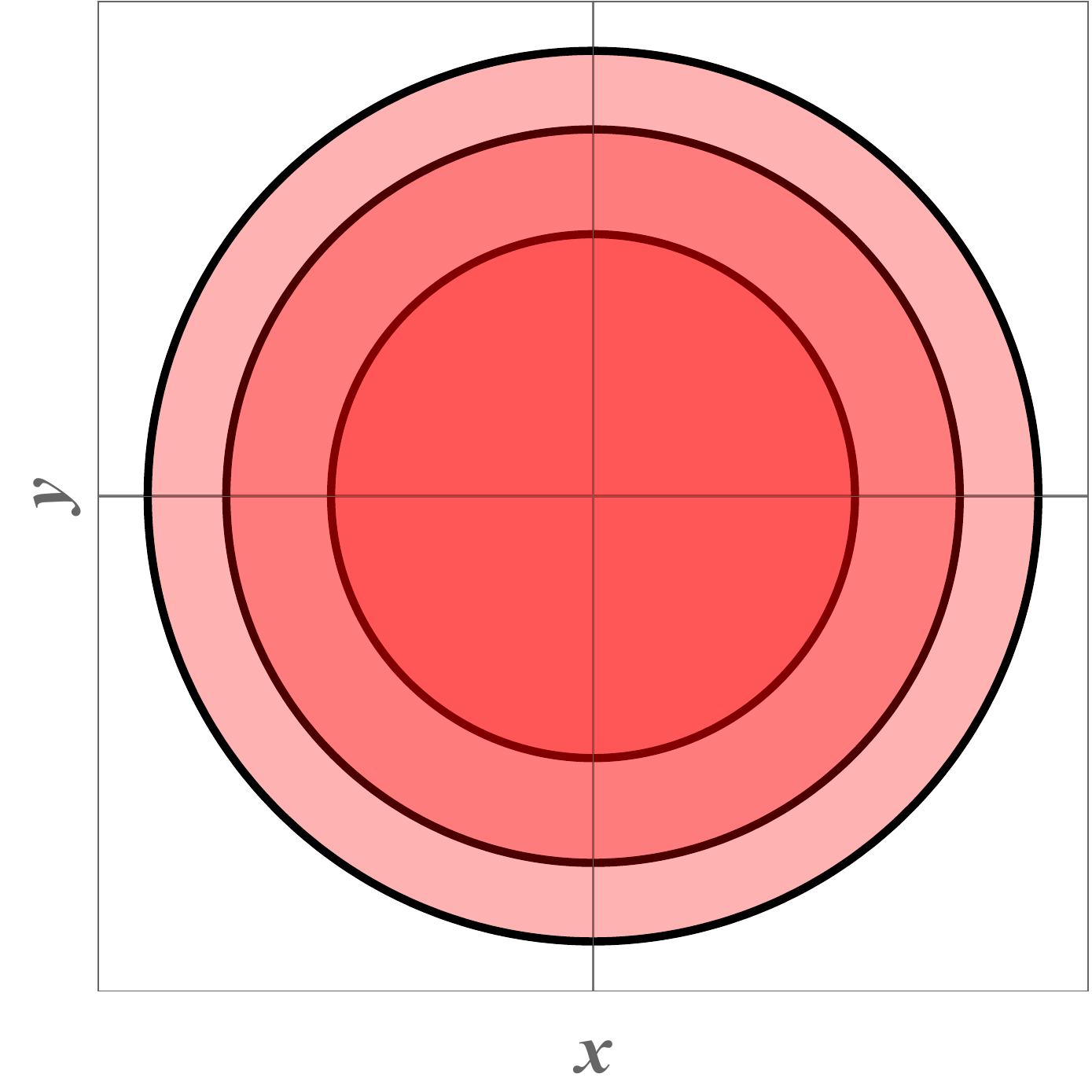}
\caption{Schematic top view of the fermions in 2d, showing the ``wedding cake'' structure of the density. In this figure there are $k_{\max}=3$ layers, with darker shaded areas corresponding to higher density.}
\label{FigWeddingCake}
\end{figure}

Let us summarize the main physical picture that emerges from our work. For
$N$ non-interacting fermions in a static non-rotating harmonic trap in
$d=2$ at zero temperature, the mean density in the limit of large $N$ takes the shape of a
spherical cap, $\rho(r)=\left(2\sqrt{2N}-r^{2}\right)/\left(4\pi\right)$, with radius $R_e \sim N^{1/4}$
and the height at the center of the cap is $\sim \sqrt{N}$ \cite{DeanEPL2015,DeanPLDReview}.
When
the rotation is switched on with a finite frequency $0<\Omega<1$ (in units such that $\omega=1$), it
still retains the shape of a spherical cap, however the rotation flattens
the density and splays the fermions over a wider region: the radius
becomes larger $R_{e}\sim\left(N/\left(1-\Omega^{2}\right)\right)^{1/4}$ while the density at the
center is smaller $\sim \sqrt{(1-\Omega^2) N}$. Finally, when the frequency
approaches its maximal permissible value for stability, $\Omega=1$, the spherical cap
approaches a flat uniform disk where all the fermions belong to the lowest
Landau level (LLL). This crossover in shape from the spherical cap to the
flat disk takes place as $\Omega$ approaches $\Omega = 1$ in a window 
$1-\Omega=O(1/N)$, and it occurs through a series of quantum steps, 
where the density takes the shape of a `wedding cake' with one uniform layer 
over another with decreasing radii (see Fig.~\ref{FigWeddingCake}). 
More precisely, 
in the window 
\be \label{mu20} 
\frac{(k_{\max}-1)k_{\max}}{N} < 1- \Omega < \frac{k_{\max}(k_{\max}+1)}{N}
\ee 
where $k_{\max}\geq 1$ is a positive integer, the total 
number of layers is $k_{\max}$, and for each layer, the density 
has a plateau value at $\rho=\frac{k}{\pi}$, $k=1,\dots,k_{\max}$. 
Finally, when $k_{\max}=1$, there is a single layer left, of density height $1/\pi$
and radius $\sim N^{1/2}$ that corresponds to the
LLL. While the wedding cake structure of the density in the vicinity of
$\Omega=1$ was found in \cite{KulkarniRotating2020}, here we study in detail the full
crossover from the finite $\Omega$ regime (spherical cap) to the uniform
disk. Our study reveals that the quantum steps also occur in the FCS and in the
entropy. More precisely, in the regime $1- \Omega=O(1/N)$, we unveil a very useful mapping 
onto a 1d harmonic oscillator, which allows to compute easily not only the steps in the density, 
but also the corresponding piecewise linear form of all the cumulants of the number of fermions in a fixed circular
domain, as well as the associated entanglement entropy. Although we restrict here to the
harmonic potential, this mapping is in principle versatile enough 
to treat more general external radial potentials such that the system
remains near the LLL, and to exhibit the universal features.

\section{Rotating harmonic oscillator in 2d: general framework} \label{sec:genf}

Let us start with the 2d geometry. The single particle Hamiltonian reads in polar coordinates $(r,\phi)$
\be
\label{eq:H2d}
H=-\frac{1}{2}\left(\partial_{r}^{2} +\frac{1}{r}\partial_{r} +\frac{1}{r^{2}}\partial_{\phi}^{2}\right)+V(r)-\Omega\frac{1}{i}\partial_{\phi} \, .
\ee 
Decomposing on the sectors of angular momentum $\ell$, i.e. the eigenfunctions are
$\psi_{n,\ell}(r,\phi)= r^{-1/2} \chi_{n,\ell}(r) \frac{e^{i \ell \phi}}{\sqrt{2 \pi}}$ with $\ell \in \mathbb{Z}$, where
the $\chi_{n,\ell}(r)$ are eigenfunctions of the radial Hamiltonian $\hat H_\ell$ with effective radial potential $V_\ell(r)$
\be \label{Vell} 
\hat H_\ell = - \frac{1}{2}  \partial_r^2 + V_\ell(r) \; , \quad 
V_\ell(r) = V(r) + \frac{\ell^2-\frac{1}{4}}{2 r^2} - \Omega \ell \; ,
\ee
see \cite{LandauLifshitz, Verbaarschot2002, KulkarniRotating2020}.

From now on we focus on the case of the harmonic potential $V(r) = \frac{r^2}{2}$,
until said otherwise below. We use units such that $\omega=1$,
$\Omega$ being now dimensionless. The eigenfunctions read
$\psi_{n,\ell}(r,\phi)= a_{n,\ell} L_n^{|\ell|}(r^2) r^{|\ell|} e^{-r^2/2} e^{i \ell \phi}$, where $L_n^{\alpha}(x)$ are the associated Laguerre polynomials, 
$a_{n,\ell}^2 = \frac{\Gamma(1+n)}{\pi \Gamma(1+ n + |\ell|)}$ and the
eigenenergies are
\be
\epsilon_{n,\ell} = 2 n+1 + |\ell| - \Omega \ell \quad , \quad n =0,1,\dots \,
\ee

\begin{figure}[h!t]
\centering
\includegraphics[angle=0,width=0.99\linewidth]{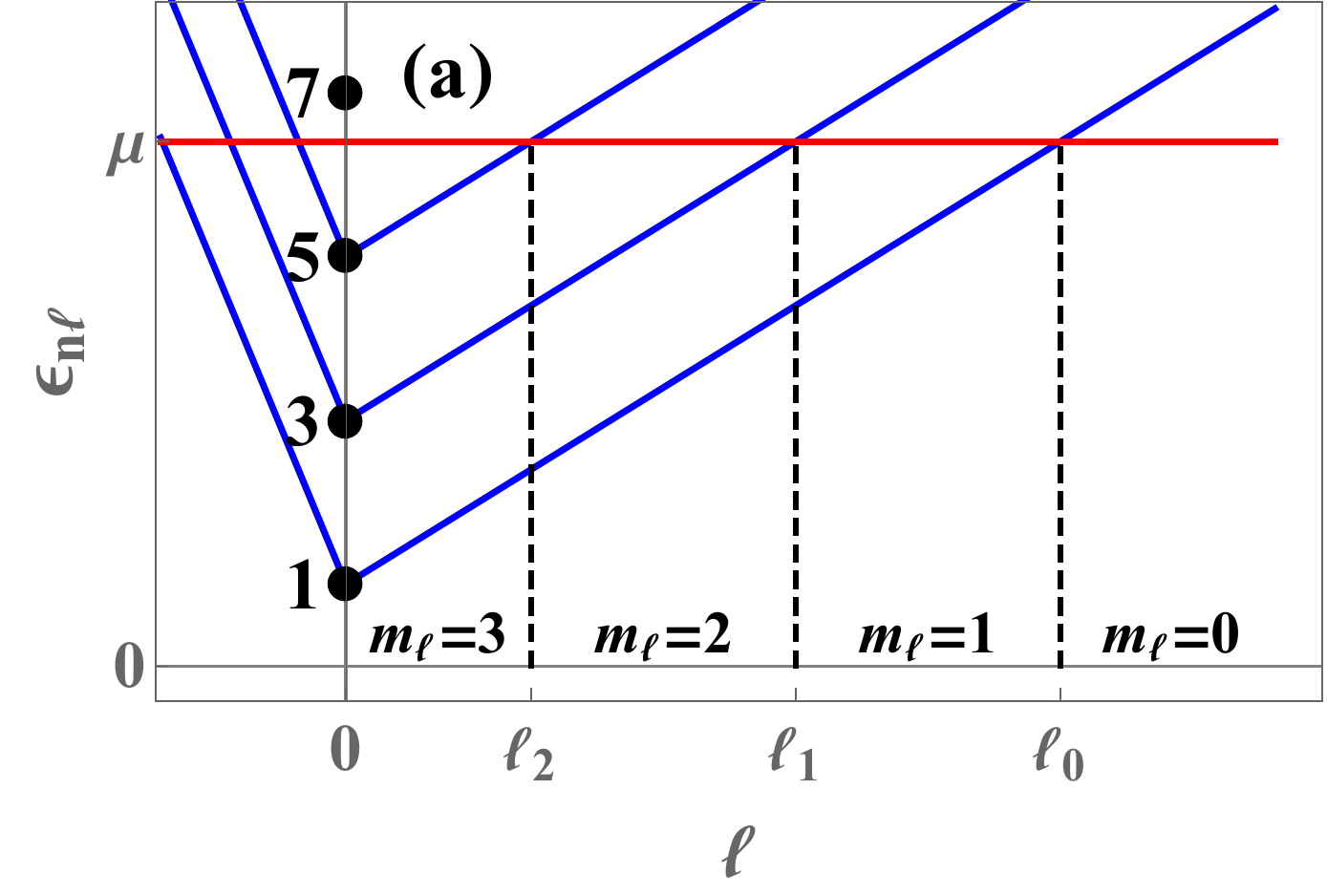}
\includegraphics[angle=0,width=0.99\linewidth]{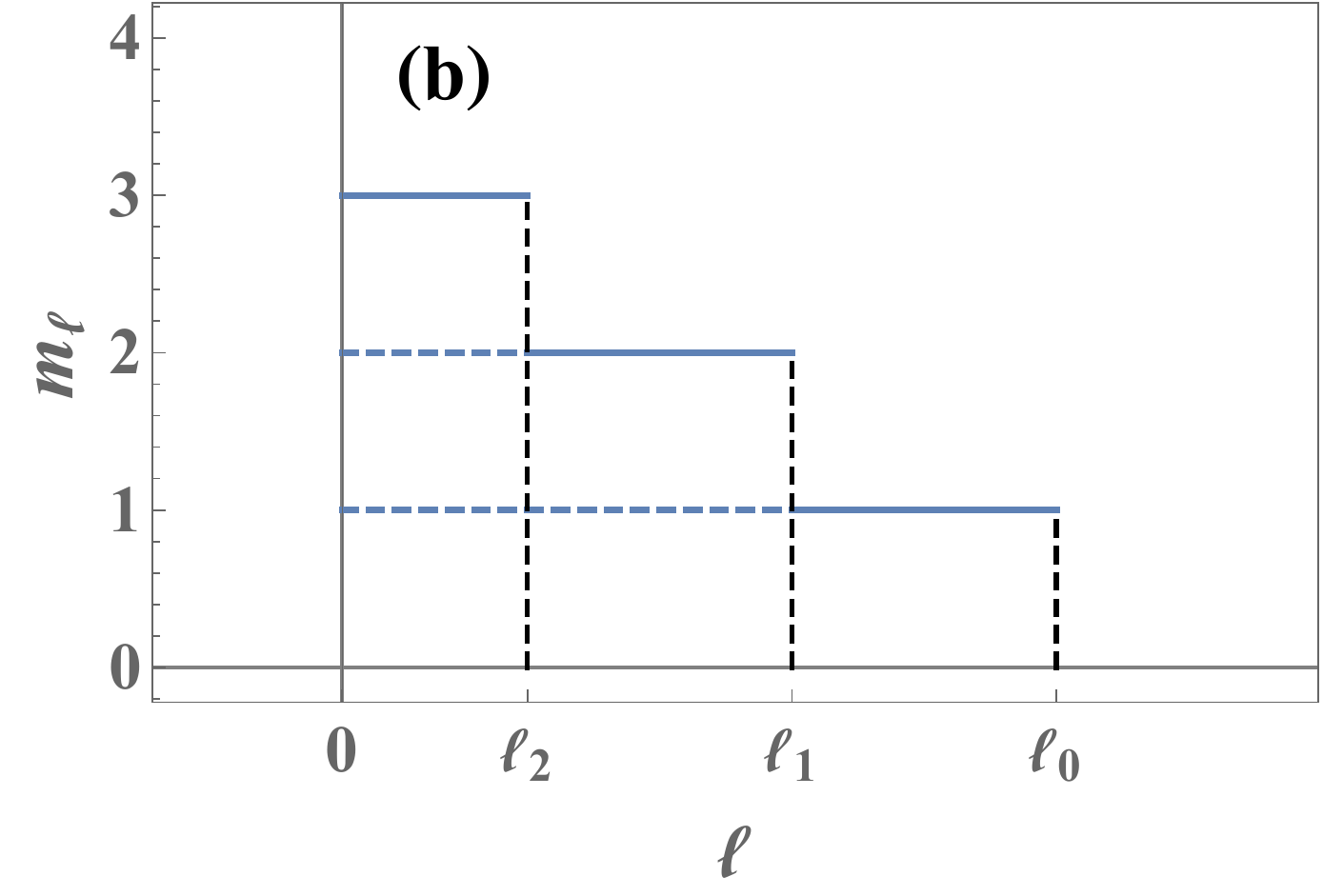}
\caption{(a) Schematic plot of the spectrum $\epsilon_{n,\ell}=2 n+1+|\ell|-\Omega \ell$ {for $0<\Omega<1$}
as a function of the angular quantum number $\ell=0,1,\dots$ for the levels $n=0,1,2,3,\dots$ {(blue solid lines, of slopes respectively $-1-\Omega$ ($\ell <0$)
and $1- \Omega$ ($\ell>0$).} 
The condition $\epsilon_{n,\ell} \le \mu$
determines the states which are occupied in the ground state. For a given $\ell$, the states with
$n=0,\dots,m_\ell-1$ are occupied. (b) Schematic plot of $m_\ell$ versus $\ell$ which exhibits jumps
at the values $\ell_k$ defined below, in Eq.~\eqref{eq:ellk} {(which are not necessarily integers).} 
}
\label{FigA}
\end{figure}

Consider now the ground state of $N$ noninteracting fermions. It is constructed as
a Slater determinant over the $N$ lowest energy eigenstates, by filling
all energy levels with $\epsilon_{n,\ell}  \leq \mu$, where $\mu$ is the Fermi energy. Let us denote by $m_\ell$ the number of
fermions in the sector $\ell$. It is equal to the number of
values of $n = 0,1,2,\dots$ such that 
$2n+1+\left[1-\Omega \, \text{sgn}\left(\ell\right)\right]\ell\leq\mu$ where $\text{sgn}\left(\ell\right)$ is the sign function. This leads to
\be
\label{eq:mell}
m_{\ell}=\begin{cases}
{\rm Int}\left(\dfrac{\mu+1-(1-\Omega)\ell}{2}\right)\quad, & \ell\geq0\\[6mm]
{\rm Int}\left(\dfrac{\mu+1+(1+\Omega)\ell}{2}\right)\quad, & \ell\leq0
\end{cases}
\ee
where ${\rm Int} (x)$ denotes the integer part of $x$.
This construction is illustrated in Fig. \ref{FigA}.
Finally the relation between $N$ and $\mu$ is given by $\sum_\ell m_\ell=N$.

Assuming $\mu \geq 1$ we see from \eqref{eq:mell} that $m_\ell \geq 1$ for $\ell_{\rm min} \leq \ell  \leq \ell_{\rm max}$
with 
\be
\label{ellmaxellmin}
\ell_{\rm max} = {\rm Int}\left( \frac{\mu-1}{1-\Omega} \right) \; , \quad \ell_{\rm min} = - {\rm Int} \left( \frac{\mu-1}{1+\Omega} \right) \; ,
\ee
and $m_\ell=0$ outside of this interval. Hence we see that when $\Omega \to 1^-$ and $\mu \geq 1$ 
many states are occupied on the
branch $\ell \geq 0$ and few on the branch $\ell <0$, see Fig.~\ref{FigA}. 
When $m_\ell=1$ for all $\ell < \ell_0$ only the LLL is filled, a situation to which we will return below.

Once the set of $m_\ell$ are known, the FCS for the number ${\cal N}_R$ of fermions in a disk of radius $R$ centered on the origin can be obtained from those of a collection of 1d problems in each angular sector. Indeed,
one can show, see e.g. \cite{UsCounting2020}, 
that the FCS generating function can be simply written as a product
\bea \label{1}
\left\langle e^{-s{\cal N}_{R}}\right\rangle =\prod_{\ell}\left\langle e^{-s{\cal N}_{[0,R]}}\right\rangle _{V_{\ell},m_{\ell}}
\eea
i.e. that the quantum fluctuations decouple in the different angular sectors. Here 
${\cal N}_{[0,R]}$ is the number of fermions in the interval $[0,R]$ for a 1d problem
of $m_\ell$ fermions in their ground state 
with Hamiltonian $\hat H_\ell$ (i.e. in the potential $V_\ell(r)$), see Eq. \eqref{Vell}. Note that 
one can explicitly write the bounds $\ell_{\rm min} \leq \ell \leq \ell_{\rm max}$ on the product \eqref{1}, or ignore it, since the factors are unity when $m_\ell=0$. In the following we analyze the
consequences of this formula in various situations. In particular taking the logarithm of
\eqref{1} and expanding in $s$ we see that each cumulant of ${\cal N}_R$ is the {\it sum} of the
cumulants of the $1d$ problems associated to the different $\ell$ sectors, i.e. for any integer
$p \geq 1$ one has \cite{footnote:pCumulant}
\be \label{cumulants0}
\left\langle {\cal N}_{R}^{p}\right\rangle ^{c}=\sum_{\ell=\ell_{{\rm min}}}^{\ell_{{\rm max}}}\left\langle {\cal N}_{[0,R]}^{p}\right\rangle _{V_{\ell},m_{\ell}}^{c}\;.
\ee

\section{Rotating harmonic oscillator in 2d: the case $0< \Omega <1$}\label{sec:3}

\subsection{FCS in the regime $0< \Omega<1$ with large $N$}
We consider the limit of large $N$, with a fixed value of the angular frequency
$0 \leq \Omega <1$, which implies large $\mu$. Let us first determine the relation between $N$ and $\mu$, as
well as the mean fermion density $\rho({\bf r})$ in the ground state, which in this problem, by rotational symmetry, is only a function of 
$r = |{\bf r}|$, i.e., 
\be 
\int d^2{\bf r}\, \rho({\bf r}) = N \; , \quad \rho({\bf r})=\rho(r) \;.
\ee 
 In that regime each angular sector has a macroscopic occupation number
$m_\ell = O(\mu)$ (see below). One can approximate $V_{\ell}(r) \simeq V_{\ell}^0(r)=\frac{r^2}{2} + \frac{\ell^2}{2 r^2}-\Omega \ell$ 
and $m_\ell \simeq \frac{\mu+ \Omega \ell - |\ell|}{2}$. We will use
the fact that for large $\mu,N$ the values of $\ell$ which dominate the
sum $N= \sum_{\ell=\ell_{\rm min}}^{\ell_{\rm max}} m_\ell$
are $\ell = O(\mu)$ with $m_\ell = O(\mu)$.
The relation between
$\mu$ and $N$ becomes 
\bea \label{rel_Nmu}
N \simeq \int_{- \frac{\mu}{1+\Omega}}^{\frac{\mu}{1-\Omega}}  d\ell \, 
\frac{\mu+ \Omega \ell - |\ell|}{2} = \frac{\mu^2}{2(1-\Omega^2)}\;. 
\eea
On the other hand the mean fermion
density in that regime is given by the semi-classical/LDA method (see Appendix \ref{sec:semiclassical} for details)
\be
\label{eq:DensityOmegaO1}
\rho({\bf r}) \simeq \frac{k_{F}(r)^{2}}{4\pi}=\frac{2\mu-(1-\Omega^{2})r^{2}}{4\pi}=\frac{2N}{\pi R_{e}^{2}}\left[1-\left(\frac{r}{R_{e}}\right)^{2}\right]
\ee 
in the disk $r<R_{e}$ where
\be \label{def_Re}
R_e=\sqrt{\frac{2\mu}{1-\Omega^{2}}}\simeq\sqrt{2}\left(\frac{2N}{1-\Omega^{2}}\right)^{1/4} \, ,
\ee
and zero outside the disk, $R>R_{e}$.
One can check using $N= \int d^2{\bf r} \rho({\bf r})$ that \eqref{rel_Nmu} is satisfied. 
 Eq.~\eqref{eq:DensityOmegaO1} shows good agreement with the numerical simulations, see Fig.~\ref{FigDensity} (a) (the technical details of the simulations are given in Appendix \ref{app:sim}).

Let us now return to the random variable ${\cal N}_R$. Its mean value is given to leading order by 
$\left\langle {\cal N}_{R}\right\rangle =\int_{r<R}d^{2}{\bf r}\rho({\bf r})\simeq N\left(\frac{R}{R_{e}}\right)^{2}\left[2-\left(\frac{R}{R_{e}}\right)^{2}\right]$,
 from \eqref{eq:DensityOmegaO1}. We now study its fluctuations, and compute the variance. 
From \eqref{cumulants0}, the variance in this regime is given by 
\bea
 {\rm Var}\,{\cal N}_{R} &=&\sum_{\ell=-\ell_{{\rm \min}}}^{\ell_{{\rm max}}}\,\left.{\rm Var}\,{\cal N}_{[0,R]}\right|_{V_{\ell},m_{\ell}} \nn\\
& \simeq & \int_{-\frac{\mu}{1+\Omega}}^{\frac{\mu}{1-\Omega}}\,d\ell\,\left.{\rm Var}\,{\cal N}_{[0,R]}\right|_{V_{\ell}^{0}(r),m_{\ell}} \; . \label{sum2} 
\eea

\begin{figure}[ht]
\centering
\includegraphics[angle=0,width=0.99\linewidth]{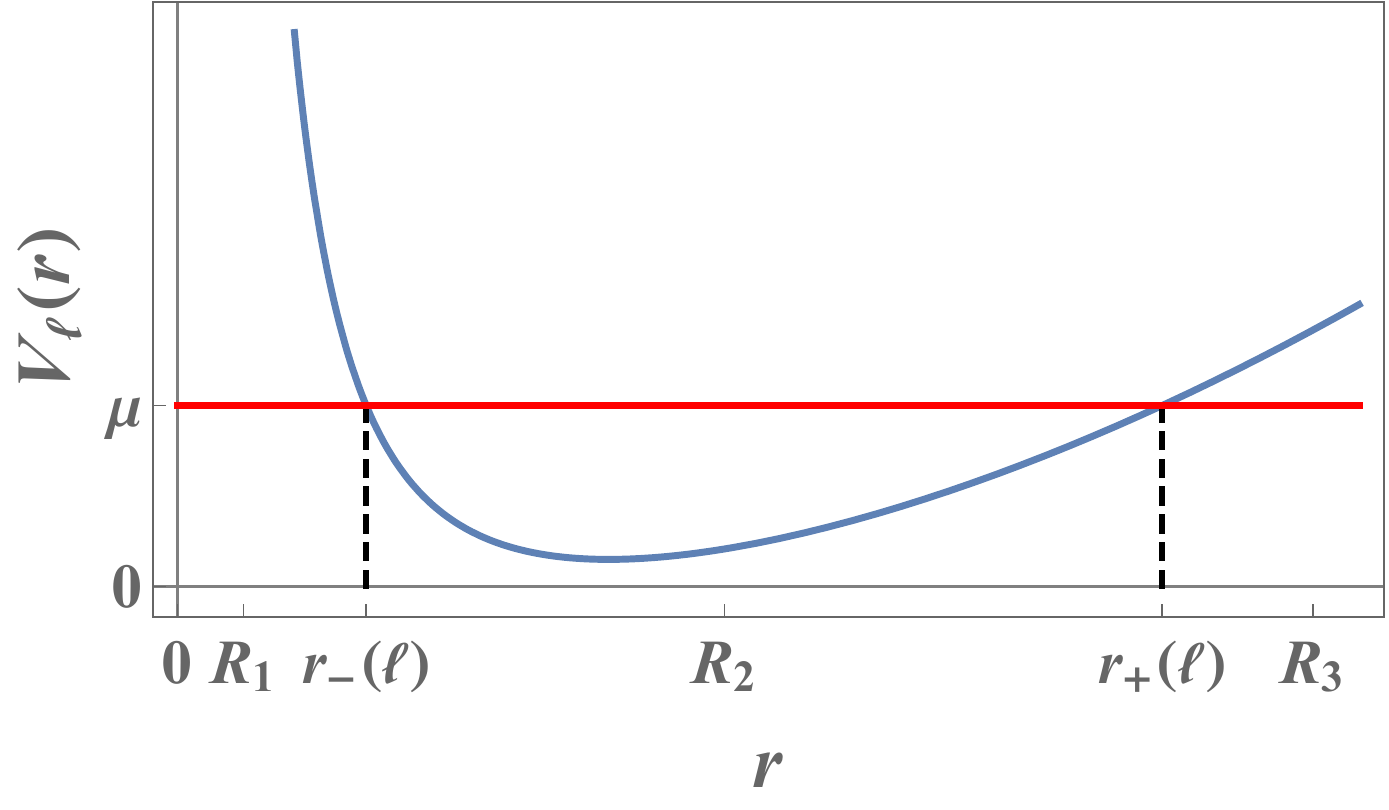}
\caption{Plot of the 1d potential $V_\ell(r)$ versus $r$ associated to a given angular sector $\ell$.
For a large number of fermions $m_\ell=O(\mu)$ in their ground state the
support of the density {$[r_-(\ell),r_+(\ell)]$} is shown. The positions of the edges $r_\pm(\ell)$ correspond to 
the roots of $V_\ell(r)=\mu$. Three cases are represented $R<r_-(\ell)$, $R \in [r_-(\ell),r_+(\ell)]$
and $R>r_+(\ell)$. In the first and last cases the fluctuations of ${\cal N}_{[0,R]}$ are negligible
since ${\cal N}_{[0,R]}\simeq 0$ or $N$ respectively.
}
\label{FigC}
\end{figure}

\begin{figure}[ht]
\centering
\includegraphics[angle=0,width=0.99\linewidth]{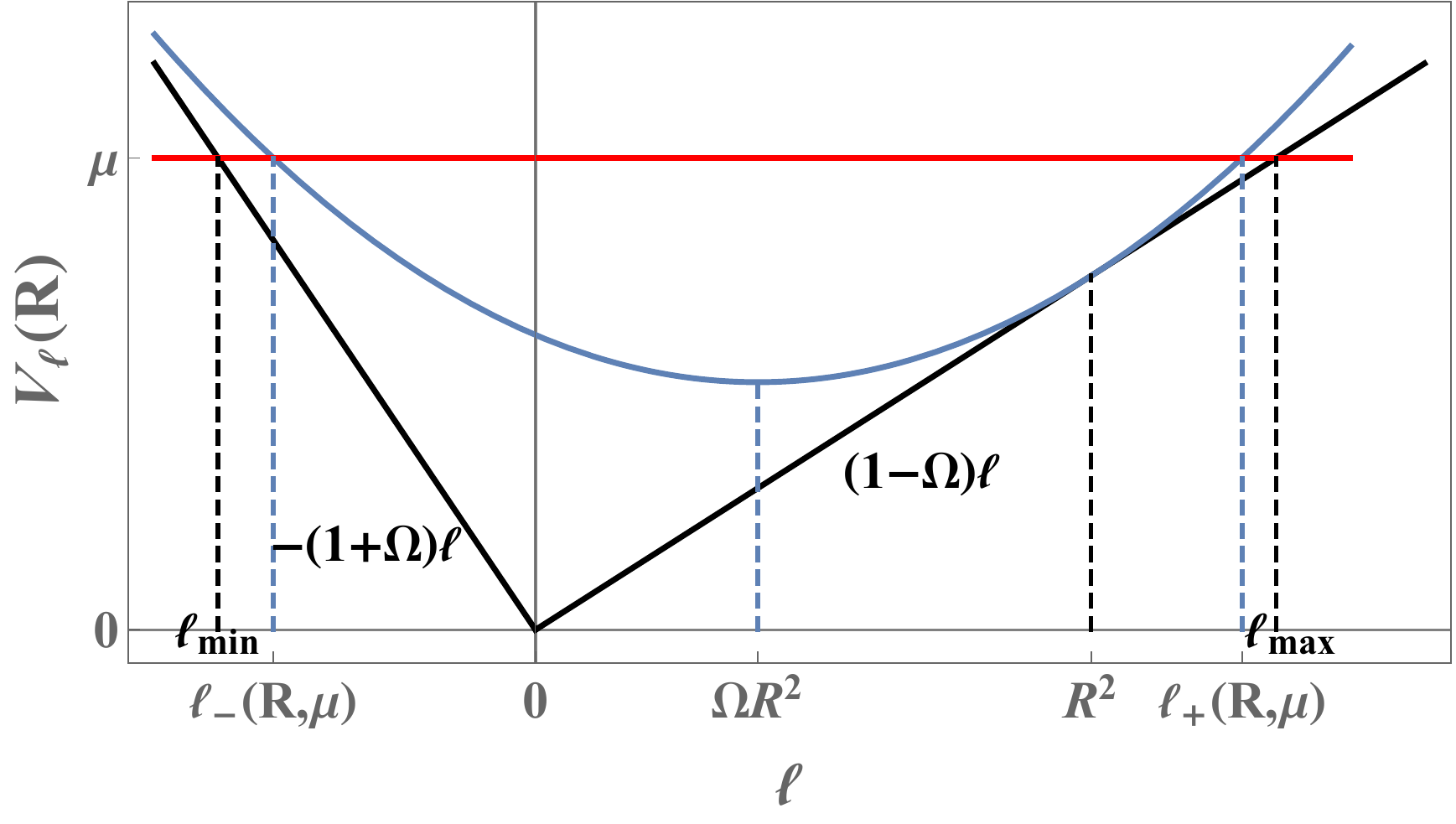}
\caption{On the same plot: (i) $\epsilon_{0,\ell}$ versus $\ell$, i.e. the two black solid lines with slopes $1-\Omega$ and $-(1+\Omega)$ respectively. Their intersections with the level $\mu$ determine $\ell_{\rm max}$ and
$\ell_{\rm min}$ defined in Eq.~\eqref{ellmaxellmin}. The distance between that level to the lines equals twice the
number $m_\ell$ of occupied states in the ground state, see Eq. (\ref{eq:mell}). 
(ii) For a given $R$, $V_{\ell}\left(R\right)  \simeq\frac{R^{2}}{2}+\frac{\ell^{2}}{2R^{2}}-\Omega\ell $ versus $\ell$ (solid blue line), is a parabola which
is tangent to both lines at $\ell=\pm R^2$. 
The intersections between $V_{\ell}\left(R\right)$ and the level 
$\mu$ determine the actual range of integration $\ell_\pm(R,\mu)$ which we use to calculate ${\rm Var}\,  {\cal N}_R$,
see Eqs.~\eqref{ellpm} and \eqref{eq:VArNRAsIntegral}.
}
\label{FigB}
\end{figure}

\begin{figure*}[ht]
\centering
\includegraphics[angle=0,width=0.48\linewidth]{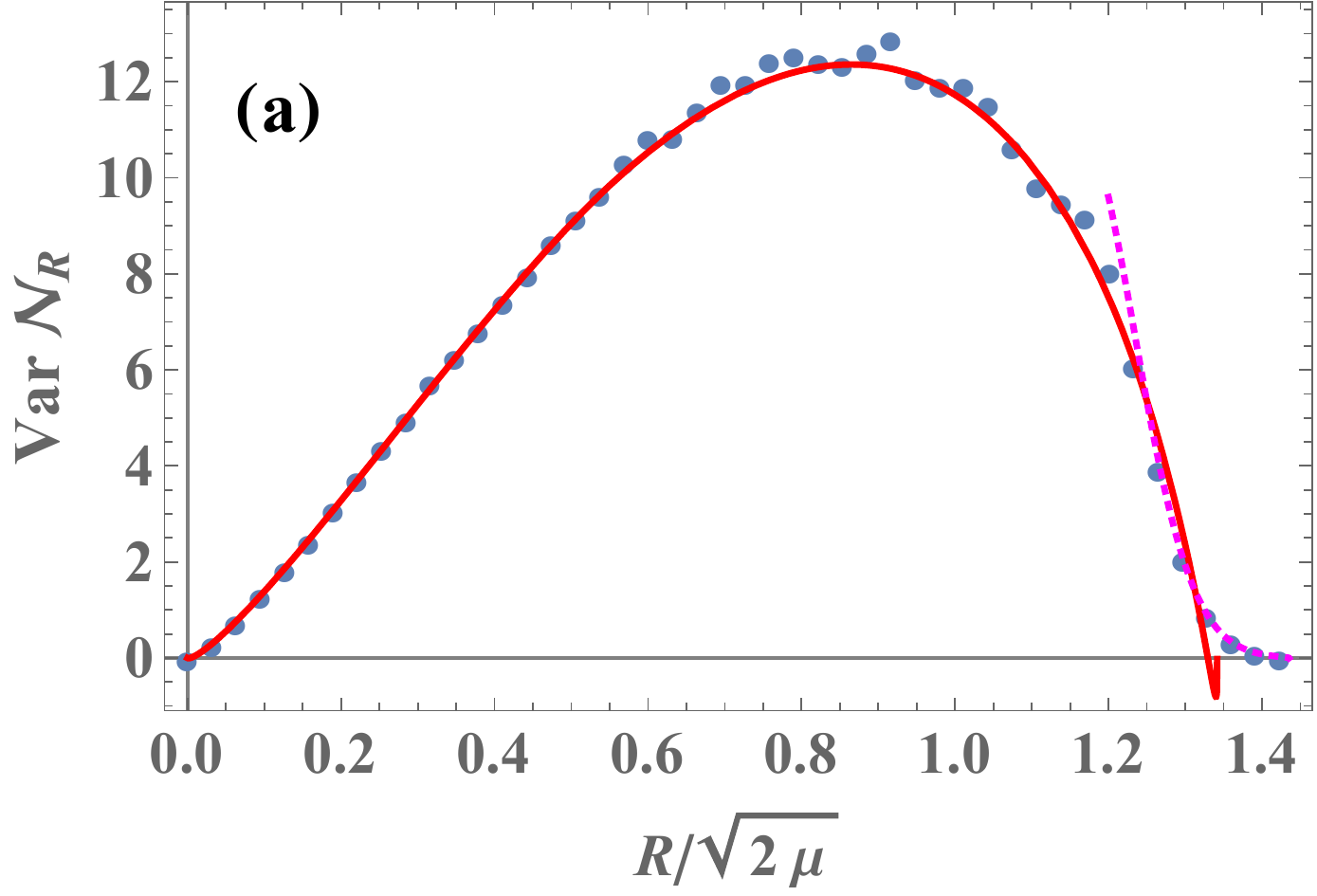}
\includegraphics[angle=0,width=0.49\linewidth]{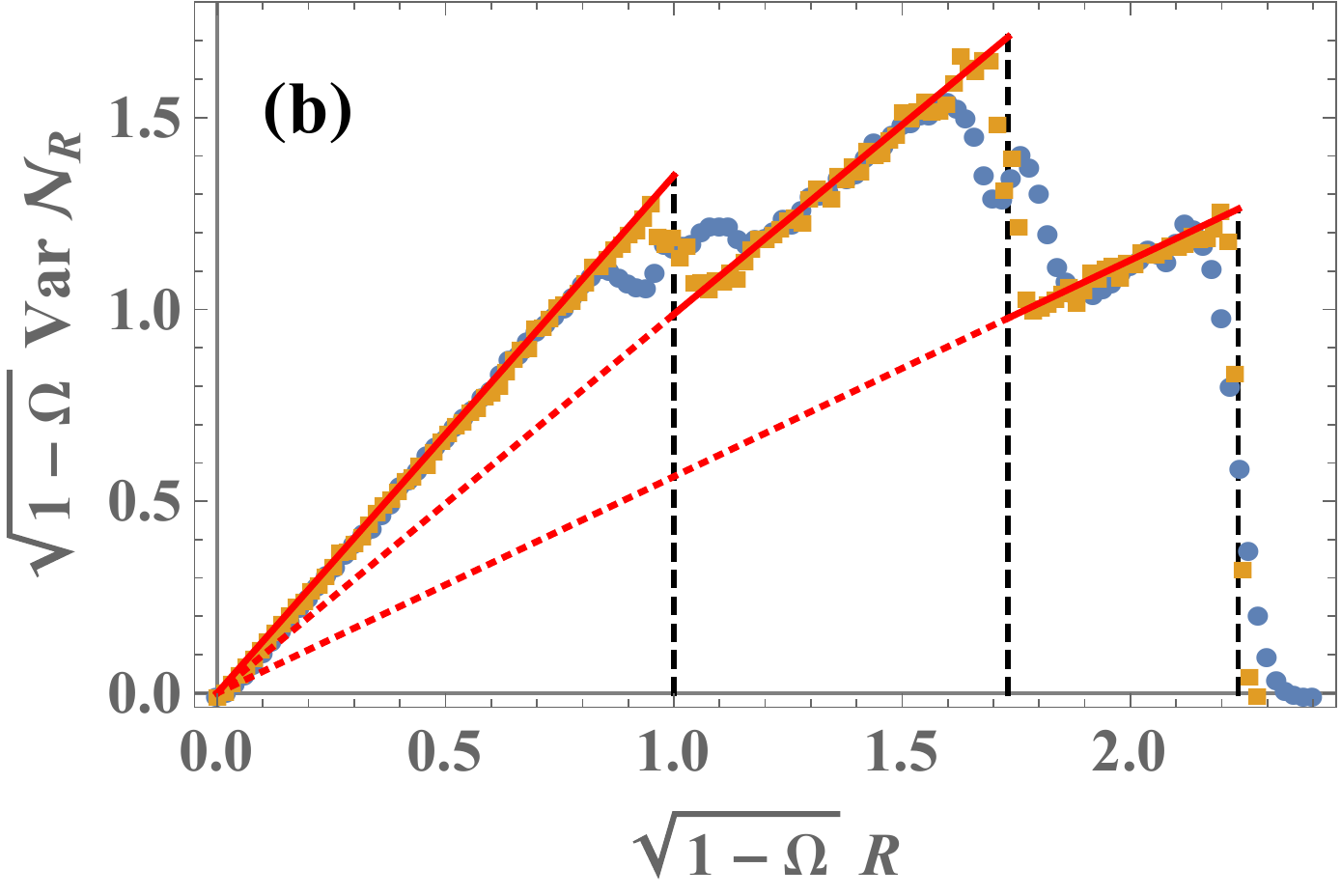}
\caption{
Variance of the number of fermions in a disk of radius $R$ for a 2d rotating harmonic trap, 
in the two regimes 
(a) $1-\Omega=O(1)$ and (b) $1-\Omega=O(1/N)$.
The symbols are the variance computed empirically over ${10}^4$ simulations. The red solid lines are the theoretical predictions \eqref{eq:VarNROmegaO1} in (a) and \eqref{eq:VarNR}  in (b), and the magenta dotted line in (a) is the edge prediction \eqref{eq:VarNREdge}. In (b), the black dashed lines correspond to $R=\sqrt{\ell_k}$ and the red dotted lines are the continuations of the formula \eqref{eq:VarNR} to the origin. Parameters are:
(a) $\Omega = 2/3$ and $\mu = 20$ corresponding to $N=366$ and
(b)  Fermi energy $\mu=6$ and $\Omega = 0.99$ (blue circles) and $\Omega = 0.999$ (yellow squares), corresponding to $N=903$ and $N=9003$ fermions respectively.
 The agreement between the theory and simulations is good except for boundary layers around $R=\sqrt{\ell_k}$ in (b), which are treated below, see, e.g., Fig.~\ref{FigVarNRWithf0Edge}. As seen in the figure, these boundary layers become narrower as $N$ is increased such that \eqref{eq:VarNR} becomes exact in the limit $N\to\infty$ limit.
}
\label{FigVarNR}
\end{figure*}

We can now use the results derived in \cite{UsCounting2020} for ${\rm Var}\,\left.{\cal N}_{[0,R]}\right|_{V_{\ell}^{0}(r),m_{\ell}}$, i.e. for the 1d inverse square potential well $V^0_\ell(r)$. In this potential
we consider $m_\ell= O(\mu)$ fermions, and the 1d fermion density 
vanishes outside the interval $[r_-(\ell),r_+(\ell)]$, where $r_\pm(\ell)$ are the two roots of
$V^0_\ell(r)=\mu$. As shown in Figure \ref{FigC}, for a fixed $\mu$ and $R$, the integrand in 
\eqref{sum2} is nonnegligible only if $R$ is inside the interval $[r_-(\ell),r_+(\ell)]$,
which is equivalent to $\ell \in [\ell_-(R),\ell_+(R)]$ where $\ell_\pm(R)$ are the two roots of 
$V^0_\ell(R) = \frac{1}{2} R^2 + \frac{\ell^2}{2 R^2}- \Omega \ell=\mu$, i.e. 
\be
\label{ellpm}
\ell_{\pm}(R) =  \Omega R^2 \pm R \sqrt{2 \mu - (1-\Omega^2) R^2} \;.
\ee
One can check that $\left[\ell_{-}(R),\ell_{+}(R)\right]\subset\left[\ell_{{\rm min}},\ell_{{\rm max}}\right]$, see Fig. \ref{FigB}.
We use the equation (18) of \cite{UsCounting2020} 
with the replacement $a \to R$, $\alpha \to \ell$, $N \to m_\ell$ and $\mu \to \mu + \Omega \ell$, leading to
\bea
\label{eq:VArNRAsIntegral}
&& 2 \pi^2 {\rm Var}\,  {\cal N}_R \simeq \\
&& \int_{\ell_{-}(R)}^{\ell_{+}(R)}d\ell\left[\log\left(R\frac{\left(2\mu-\left(R^{2}+\frac{\ell^{2}}{R^{2}}-2\Omega\ell\right)\right)_{+}^{3/2}}{\left((\mu+\Omega\ell)^{2}-\ell^{2}\right)_{+}^{1/2}}\right)+c_{2}\right] \nn
\eea 
with $c_2  =  \gamma_E +1+\log 2$, where $\gamma_E$ is Euler's constant.
To be accurate we must stress that this formula holds in the following sense.
Define the scaled radius $\tilde R = R/\sqrt{2 \mu}$. Then in the limit of large $\mu$ with 
fixed $\tilde R$ and $\Omega$, we obtain our main result for the variance
\bea
\label{eq:VarNROmegaO1}
&&   {\rm Var}\,  {\cal N}_R = A_{\tilde R,\Omega} \mu \log \mu + B_{\tilde R,\Omega}  \mu + o(\mu) \, , \\
&& A_{\tilde R,\Omega}  = \frac{2}{\pi^2} \tilde R \sqrt{1 - (1- \Omega^2) \tilde R^2} \, , \label{eq:AROmega} \\
\label{eq:BROmega}
&& B_{\tilde{R},\Omega}=\int_{\lambda_{-}(\tilde{R})}^{\lambda_{+}(\tilde{R})}\frac{d\lambda}{2\pi^{2}} \nn\\
&& \quad \times \left[\log\left(4\tilde{R}\frac{(1-(\tilde{R}^{2}+\frac{\lambda^{2}}{4\tilde{R}^{2}}-\Omega\lambda))_{+}^{\frac{3}{2}}}{((1+\Omega\lambda)^{2}-\lambda^{2})_{+}^{\frac{1}{2}}}\right)+c_{2}\right]  ,
\eea 
where $\lambda_{\pm}=2\tilde{R}\left(\Omega\tilde{R}\pm\sqrt{1-(1-\Omega^{2})\tilde{R}^{2}}\right)$.
The integral in the formula for $B_{\tilde{R},\Omega}$ can be performed and is displayed in \eqref{eq:BIntegralExplicit}.
Note that in the scaled variables the position of the edge is $\tilde R_e= \frac{1}{\sqrt{1- \Omega^2}}>1$
(the rotation expands the Fermi gas). Alternatively one can express \eqref{eq:VarNROmegaO1}-\eqref{eq:BROmega} in terms of $N$
and $R/R_e$ using \eqref{rel_Nmu} and $\tilde R=R/(R_e \sqrt{1-\Omega^2})$. 
In Fig.~\ref{FigVarNR} (a), the numerical simulations of ${\rm Var}\,  {\cal N}_R$ are compared with the prediction of \eqref{eq:VarNROmegaO1}, showing excellent agreement.

It is also interesting to study the $p^{\rm th}$ cumulant of ${\cal N}_R$, i.e. $\left\langle {\cal N}_{R}^{p}\right\rangle ^{c}$,
beyond the variance, which corresponds to $p=2$. By the same small scale universality arguments as in Ref. \cite{UsCounting2020}, we obtain (for arbitrary integer $m>1$)
\be \label{cumulants}
\left\langle {\cal N}_{R}^{2m}\right\rangle ^{c}=2\mu\tilde{R}\sqrt{1-(1-\Omega^{2})\tilde{R}^{2}}\,\left(\kappa_{2m}+o(\mu)\right)\;,
\ee
with 
\be
\label{kappDef}
\kappa_{2m} = {(-1)^{m+1}(2m)! 2 \zeta(2m-1)}/{(m (2 \pi)^{2m})}
\ee
 while the odd cumulants vanish to $O(\mu)$.

\subsection{Edge regime for $ 1-\Omega = O(1)$}

We now calculate the number variance for $R\simeq R_e$ near the edge of the Fermi gas in the regime $1-\Omega=O(1)$.
Our result \eqref{eq:VarNROmegaO1} breaks down near the edge, because it was obtained using approximations that are only valid in the bulk.

For $1d$ systems, the statistical properties near the edge of Fermi gases are determined by the Airy kernel
\be
K_{\rm Ai}(x,y)=\frac{{\rm Ai}(x) {\rm Ai}'(y)-  {\rm Ai}'(x) {\rm Ai}(y)}{x-y} \, .
\ee
In particular, the number variance for a semi-infinite interval beginning (or ending) around the edge of the gas follows a known universal behavior which depends only on the derivative of the potential at the edge \cite{MMSV14, MMSV16}. Applying these known results to each of the sectors, we have
\be
\label{eq:VarSectorsV2}
\left.{\rm Var}\,{\cal N}_{[0,R]}\right|_{V_{\ell}^{0}(r),m_{\ell}}\simeq\frac{1}{2}\mathcal{V}_{2}\left(\frac{R-r_{+}\left(\ell\right)}{w\left(\ell\right)}\right)
\ee
where $w\left(\ell\right)=\left[2\left(V_{\ell}^{0}\right)'\left(r_{+}\right)\right]^{-1/3}$ and the universal scaling function is \cite{MMSV14, MMSV16}
\be
{\cal V}_2(\hat a) := {2}
\int_{\hat a}^{+\infty}du\int_{-\infty}^{\hat a}dv K_{\text{Ai}}^{2}\left(u,v\right) \, .
\ee
These results are valid when the argument of ${\cal V}_2$ in \eqref{eq:VarSectorsV2} is of order unity.

We now wish to use \eqref{eq:VarSectorsV2} in \eqref{sum2} in the regime $R \simeq R_e$. The dominant contribution to ${\rm Var}\,{\cal N}_{R}$ is from $\ell \simeq \ell_e$ where $r_{+}\left(\ell_e\right)=R_e$ and therefore, plugging \eqref{eq:VarSectorsV2} directly into \eqref{sum2} gives the correct leading-order result. Using \eqref{ellpm} with $R=R_e= \sqrt{\frac{2 \mu}{1-\Omega^2}}$, we find that $\ell_{e}= \Omega R_e^2=2\mu\Omega/\left(1-\Omega^{2}\right)$. At $\ell \simeq \ell_e$, 
\bea
\!\!\! w\left(\ell\right) &\simeq& w\left(\ell_{e}\right)=\left[2\left(V_{\ell_{e}}^{0}\right)'\left(R_{e}\right)\right]^{-1/3} \nn\\
&=& \left[2\left(R_{e}-\frac{\ell_{e}^{2}}{R_{e}^{3}}\right)\right]^{-1/3} \! =\left[8\mu\left(1-\Omega^{2}\right)\right]^{-1/6}  .
\eea
Summarizing, we have so far
\be
\label{eq:VarSectorsV2a}
{\rm Var}\,{\cal N}_{R}\simeq\int_{-\frac{\mu}{1+\Omega}}^{\frac{\mu}{1-\Omega}}\,d\ell\,\frac{1}{2}\mathcal{V}_{2}\left(\frac{R-r_{+}\left(\ell\right)}{w\left(\ell_{e}\right)}\right) \, .
\ee
Solving the equation $V_{\ell}^{0}\left(r_+\right)=\mu$, we find
\be
r_{+}^{2}=\left(\Omega\ell+\mu\right)+\sqrt{\left(\Omega\ell+\mu\right)^{2}-\ell^{2}}
\ee
which, in the vicinity of $\ell=\ell_e$ (and $r_+ = R_e$), yields
\be
\label{rplusApprox}
r_{+}\simeq R_{e}-\frac{\sqrt{1-\Omega^{2}}}{4\sqrt{2}\,\mu^{3/2}}\left(\ell-\ell_{e}\right)^{2} \, .
\ee
Plugging \eqref{rplusApprox} into \eqref{eq:VarSectorsV2a}, we find that the dominant contribution to the integral comes from $\ell \simeq \ell_e$ and therefore the integration limits can be replaced by plus and minus infinity:
\be
{\rm Var} \, {\cal N}_{R}\simeq\int_{-\infty}^{\infty}\!\!d\ell\,\frac{1}{2}\mathcal{V}_{2}\left(\frac{R-R_{e}}{w\left(\ell_{e}\right)}+\frac{\sqrt{1-\Omega^{2}}\,\left(\ell-\ell_{e}\right)^{2}}{4\sqrt{2}\,\mu^{3/2}w\left(\ell_{e}\right)}\right)\,.
\ee
Now, we change the integration variable to $\xi=\frac{\sqrt{1-\Omega^{2}}}{4\sqrt{2}\,w\left(\ell_{e}\right)\mu^{3/2}}\left(\ell-\ell_{e}\right)^{2}$,
noting that the two sides $\ell > \ell_e$ and $\ell < \ell_e$ produce identical results, leading to an extra factor of $2$. Simplifying, we
obtain finally the scaling form for the variance at the edge in the regime $\Omega<1$
\be
\label{eq:VarNREdge}
{\rm Var}\,{\cal N}_{R}\simeq\frac{\mu^{2/3}}{\left(1-\Omega^{2}\right)^{1/3}}\int_{0}^{\infty}\,\frac{d\xi}{\sqrt{\xi}}\mathcal{V}_{2}\left(\hat{R}+\xi\right)
\ee
where $\hat{R} = \frac{R-R_{e}}{w\left(\ell_{e}\right)}$. In the nonrotating case $\Omega=0$, Eq.~\eqref{eq:VarNREdge} coincides with the result of \cite{UsCounting2020}.

\subsection{Entanglement entropy in the bulk}

We now apply our results to the calculation of the bipartite R\'enyi entanglement entropy of a disk $D_R$ of radius $R$ centered around the origin with its complement $\overline{D}_R$. 
For any domain $\mathcal{D}$, its definition, parametrized by $q  \geq  1$, is given by
\be
S_q({\cal D}) = \frac{1}{1-q}\log \Tr[\hat{\rho}_{\cal D}^q],
\ee
 where the reduced density matrix is given by $\hat{\rho}_{\cal D}  =  \Tr_{\overline{\cal D}}[\hat{\rho}]$, i.e., tracing out the density matrix $\hat{\rho}$  of the system over $\overline{\cal D}$.
In the limit $q \to 1$ one recovers the von Neumann entropy $S_{1}=-\text{Tr}\left[\hat{\rho}_{{\cal D}}\log\hat{\rho}_{{\cal D}}\right]$.
A remarkable fact is that for noninteracting fermions one can express
$S_q({\cal D})$ as a series  in the (even) cumulants of ${\cal N}_{\cal D}$, 
\be
\label{SqSumCumulants}
S_{q}({\cal D})=\sum_{m\geq1}s_{m}^{(q)}\left\langle {\cal N}_{{\cal D}}^{2m}\right\rangle ^{c},
\ee
where the coefficients  $s_m^{(q)}$ are given in \cite{CMV2012}
and $s_1^{(q)} =  \frac{\pi^2}{6}(1+ \frac{1}{q})$. 

For a disk of radius $R$, in the regime $\Omega = O(1)$ we use our expressions \eqref{eq:VarNROmegaO1} for the variance  and \eqref{cumulants} for the higher cumulants to obtain
\bea
\label{entropy}
S_{q}\left(D_{R}\right)&=&\frac{\pi^{2}\left(q+1\right)}{6q}{\rm Var}\,{\cal N}_{R} \nn\\
&+&2\mu\tilde{R}\sqrt{1-(1-\Omega^{2})\tilde{R}^{2}}\,\tilde{E}_{q}+o(\mu)
\eea
where $\tilde E_q=\sum_{m\geq2}s_{m}^{(q)}\kappa_{2m} =  E_q   -  \frac{q+1}{6 q} (1 + \gamma_E)$ and
 $E_{q}$ is given explicitly in Eq.~(11) in \cite{CalabreseEntropyFreeFermions} (see also \cite{JinKorepin2004}).
In \eqref{entropy} the simple form of the second term 
arises from the common $R$ dependence of the 
cumulants of order $4$ and higher.
For the nonrotating case $\Omega = 0$, the result \eqref{entropy} is in agreement with \cite{UsCounting2020}.

Note that in the other regime $\Omega \to 1$, the entanglement entropy in the lowest Landau level $k=1$ was 
obtained directly from the overlap matrix in \cite{LMG19}. Since all the cumulants are linear in $R$,
the entropy was also found to be linear in $R$ in the bulk (see also Refs. \cite{RS09,CE19,LSS20} for studies of the entanglement entropy in that regime). We will return to this point in the next section.

\section{Rotating harmonic oscillator in 2d: the case $1-\Omega \to 0$} \label{sec:LLL}

 We now address the limit $\Omega \to 1^-$ and, in the large $N$ limit, the regime $1-\Omega=O(1/N)$.
As described in the Introduction, a sequence of transitions occurs in this regime, corresponding to filling 
successive Landau levels, at discrete values of $\Omega$ given by 
\eqref{mu20} (see also \eqref{mu2} below with $n=k_{\max}$). We first address the case of a single LLL, where the density
is uniform in a disk, and in a second stage we decrease $\Omega$, 
which results in higher Landau levels being occupied and
new layers being formed, see Fig.~\ref{FigWeddingCake}.

\subsection{LLL and FCS in the regime $\Omega \to 1^-$}

Let us first recall the case $1- \frac{2}{N-1} < \Omega < 1$ where the ground state is constructed from single particle states in the LLL
(i.e. the branch $n=0$ in Fig.~\ref{FigA}. This case is related to the Ginibre ensemble of random matrices \cite{LMG19}. 
Here, as a benchmark for our method, we show how to recover the results of \cite{LMG19} using a 
different eigenbasis for single particle states. From Fig.~\ref{FigA} we see that the occupied energy levels in the ground state
are $\epsilon_{0,\ell}= 1 + (1-\Omega) \ell$ for $\ell=0,\dots,N-1$, i.e. no states on the branch $\ell<0$ 
are occupied and the occupied states on the branch $\ell\geq 0$ all contain a single fermion, $m_\ell=1$. 
For a single fermion in an inverse square potential it is easy to obtain the probability that 
the fermion is in $[0,R]$ from its wave function $\psi_{0,\ell}$. From \eqref{1} one then immediately obtains the FCS generating function as
\bea \label{1LLL0}
\left\langle e^{-s{\cal N}_{R}}\right\rangle  &=&\prod_{\ell=0}^{N}\left\langle e^{-s{\cal N}_{[0,R]}}\right\rangle _{V_{\ell},m_{\ell}=1} \\
&=& \prod_{\ell=0}^{N}\int_{0}^{+\infty}\left[1-\left(1-e^{-s}\right)\int_{0}^{R}\chi_{0\ell}(r)^{2}dr\right] \nn
\eea
with $\chi_{0\ell}(r)=\sqrt{\frac{2}{|\ell| !}} r^{|\ell|+ \frac{1}{2}} e^{-r^2/2}$.
Taking the logarithm one finds the cumulants 
\be
\sum_{p=0}\!\frac{(-s)^{p}}{p!}\!\left\langle {\cal N}_{R}^{p}\right\rangle ^{c}\!=\!\sum_{\ell=0}^{N-1}\!\log\!\left(1-(1-e^{-s})\frac{\gamma\left(\ell+1,R^{2}\right)}{\Gamma(\ell+1)}\right)
\ee 
where $\gamma(a,z)=\int_0^z t^{a-1} e^{-t} dt$, recovering Eq. (31) in {the supplemental material of \cite{LMG19}}.  

These results are exact for any $N$. In the large $N$ limit this corresponds 
to taking the limit $\Omega \to 1^-$ with $\Omega$ in the window $1- \frac{2}{N} < \Omega < 1$.
In that regime, the analysis of \eqref{1LLL0} performed in \cite{LMG19} shows that the
cumulants grow linearly with the radius of the disk $R$ as
\be
\left\langle {\cal N}_{R}^{p}\right\rangle ^{c}\simeq B_{1}^{(p)}R
\ee 
where the $B^{(p)}_1$ are numbers obtained in \cite{LMG19} 
with $B^{(2)}_1 = {1}/{\sqrt{\pi}}$.

\subsection{Vicinity of the LLL and $\Omega \to 1$ limit}

We now explore the vicinity of the LLL, focusing on the large $N$ limit 
with $1-\Omega = O(1/N)$. We unveil a mapping to a 1d harmonic
oscillator, Eq. \eqref{Vell2} below, which allows to easily compute the steps in the density
(the "wedding cake") and to obtain the corresponding quantum step structure for the FCS. 

We first discuss the filling of the Landau levels. It is equivalent to decrease $\Omega$ at fixed $N$
or to increase the Fermi energy $\mu$ for a fixed $\Omega$.
Let us focus first on the latter and denote by $n \ge 0$ the integer for which
\be \label{mu} 
2 n+1 < \mu < 2 n+3 \,.
\ee 
Consider the energy levels with $\ell \geq 0$.
The level $\epsilon_{n,\ell}$ is occupied if $2 n+1 + (1-\Omega)\ell \leq \mu$. Hence we find that
$m_\ell$ is a decreasing integer staircase (see figure \ref{FigA}), with $m_0=n+1=m_\ell$ for $\ell<\ell_{n}=\frac{\mu-(2n+1)}{1-\Omega}$, and $m_\ell=0$ for
$\ell > \ell_0=\frac{\mu-1}{1-\Omega}$, more generally
\bea
\label{eq:ellk}
&& m_\ell = k \quad , \quad \ell_k < \ell < \ell_{k-1} \quad , \quad k=n+1,\dots,0 \nn \\
&& \ell_k = \frac{\mu - (2k+1)}{1- \Omega} 
\eea
where the $\ell_k$ are not necessarily integers.
The LLL case is obtained for $n=0$, i.e. $1 < \mu < 3$, in which case $m_\ell =1$
for $\ell < \ell_0=\frac{\mu-1}{1-\Omega}$. Since we want the states with $\ell=0,\dots,N-1$ to be occupied
we need $\ell_0 > N-1$ hence $1-\Omega <(\mu-1)/(N-1)$, which for large $N$
leads to the window $1- \frac{2}{N} < \Omega<1$ identified above and studied in \cite{LMG19}.
As we show below the cases $n=1,2,\dots$ lead to a structure similar, but slightly different to the "wedding cake" of \cite{KulkarniRotating2020}, where the number of layers decreases as $\ell$ increases. 
To compute the relation between $\mu$ and $N$ we note that the total number
$N_{\geq}$ of occupied states for $\ell \geq 0$ is
\bea
N_{\geq} &\simeq& (n+1) \ell_n + n (\ell_{n+1}-\ell_{n}) + \dots \nn\\
&=&
\sum_{k=0}^{n} \ell_k = \frac{(n+1) (n + \mu')}{1-\Omega} 
\eea 
with $\mu'=\mu-(2n+1) \in ]0,2[$ [see Eq. (\ref{mu})]. Since in that regime the number $N_{<}$ of
occupied states with $\ell<0$ is only $O(1)$ it can be neglected so $N \simeq N_{\geq}$
hence the regime \eqref{mu} is equivalent to 
\be \label{mu2} 
\frac{n(n+1)}{N} < 1- \Omega < \frac{(n+1)(n+2)}{N}
\ee 
This is the large $N$ regime on which we focus here. In that case $\mu= O(1)$ 
and the size of Fermi gas is of order $R_e \sim \sqrt{N}$, see below.
\\

We now discuss the statistics of ${\cal N}_R$ in that regime, i.e. described by \eqref{mu}, \eqref{mu2}
with $n$ a fixed integer and $N$ large. A first simplification
is that it can be mapped to a 1d harmonic oscillator problem. 
Indeed, we see that the $\ell_k$ are typically $O(N)$.
One can check that for large $\ell$ the potential $V_\ell(r)$ in \eqref{Vell} is very well approximated around its minimum 
at $r=r^*(\ell)$ by a quadratic potential. In the present case of the harmonic oscillator (HO), $V(r) = \frac{1}{2} r^2$
one finds $r^*(\ell)=\sqrt{\ell}$ and denoting $r = r^*(\ell) + u$ one obtains the expansion
\be \label{Vell2} 
V_{\ell}\left(r=\sqrt{\ell}+u\right)\simeq\left(1-\Omega\right)\ell+2u^{2}+O\left(u^{3}/\sqrt{\ell},u^{4}/\ell\right) \, .
\ee
Thus we can approximate the original Hamiltonian $H_\ell$ in each angular sector by a
harmonic oscillator with Hamiltonian $H_{1d\rm{HO}} = - \frac{1}{2} \frac{\partial^2}{\partial u^2} + 2 u^2$.
As a consequence [see Eq. (\ref{Vell2})] we have 
$\left.{\cal N}_{[0,R]}\right|_{V_{\ell}(r),m_{\ell}}\simeq\left.{\cal N}_{\left[-\sqrt{\ell},R-\sqrt{\ell}\right]}\right|_{2u^{2},m_{\ell}}$,
where the subscripts in the last quantity indicated that it is calculated in the ground state of $H_{1d\rm{HO}}$ with $m_\ell$ fermions.

Let us begin by calculating the density $\rho\left({\bf r}\right) \equiv \rho\left(r\right)$, which is related to the mean of ${\cal N}_{R}$ via $\frac{d\left\langle {\cal N}_{R}\right\rangle }{dR}\big \vert_{R=r}=2\pi r\rho\left(r\right)$. We consider here $R=O(\sqrt{N})$, i.e. $\tilde R=O(1)$. 
Denoting by $\rho_{1d\text{HO},k}\left(u\right)$ the mean density for $k$ noninteracting fermions in the ground 
state of $H_{1d\rm{HO}}$ one has
\bea \label{density1} 
\left\langle {\cal N}_{R}\right\rangle &\simeq& \sum_{k=1}^{n+1}\int_{\ell_{k}}^{\ell_{k-1}}d\ell\,\left\langle \left.{\cal N}_{\left[-\sqrt{\ell},R-\sqrt{\ell}\right]}\right|_{2u^{2},k}\right\rangle \, , \\
2\pi r\rho\left(r\right) &\simeq& \sum_{k=1}^{n+1}\int_{\ell_{k}}^{\ell_{k-1}}d\ell\,\rho_{1d\text{HO},k}\left(r-\sqrt{\ell}\right) \, . 
\label{eq:DensityOmegaNearOne} 
\eea 
We note now that for each value of $r$ only one value of $k$ contributes. Indeed
the sizes of the staircase steps, $\sqrt{\ell_{k-1}}-\sqrt{\ell_{k}}$ being much larger than the width of the ground-state wavefunction of the harmonic oscillator (which is $O(1)$), for each $r$ the equivalent HO problem has a fixed number of fermions $k$
determined by $\sqrt{\ell_k} < r < \sqrt{\ell_{k-1}}$. Hence we have
\bea 
&& 2\pi r\rho\left(r\right) \label{dens22} \\
&&\simeq2 r\,\theta\left(\sqrt{\ell_{k}}<r<\sqrt{\ell_{k-1}}\right) \int_{-\infty}^{+\infty}du \, \rho_{1d\text{HO},k}\left(u\right) \nn\\
&& =2kr\quad ,\quad\sqrt{\ell_{k}}<r<\sqrt{\ell_{k-1}} \;. \label{densfinal} 
\eea
Since the width of the equivalent oscillator is fixed, to obtain \eqref{dens22} we
changed the integration variable
as $\sqrt{\ell}=r - u$ where $u$ varies of $O(1)$, and note that $d\ell=-2\left(r-u\right)du\simeq-2r \,du$.
{
Finally, to reach Eq.~\eqref{densfinal} we used the normalization of the density for the 1d HO, $\int_{-\infty}^{+\infty}\rho_{1d\text{HO},k}\left(u\right)du=k$.
Our prediction \eqref{densfinal} 
for the density is compared with numerical simulations in Fig.~\ref{FigDensity} with excellent agreement.
Note that the index $n+1$ in \eqref{mu2} gives the total number of Landau level which are occupied,
while the index $k=1,\dots,n+1$ labels the plateaus in the density at $\rho(r)=k/\pi$,
and correspond to a ``local" droplet of a $k$-th Landau level. The density 
\eqref{densfinal}, describing a ``wedding cake'' structure, i.e. with plateaux at
$\rho(r)= \frac{k}{\pi}$, was obtained in \cite{KulkarniRotating2020} in a more general setting \cite{footnote:density}.
They arrived at this result by first calculating the exact density and then performing an asymptotic analysis in the large-$N$ limit. 
They also studied the boundary layer form of the density around the edges of each plateau, at $r = \sqrt{\ell_k}$ where \eqref{densfinal} breaks down. In Appendix \ref{sec:stepStructureRho} we compute directly this
boundary layer form using the harmonic approximation \eqref{Vell2} and show that it is adequate in order to recover
the result of \cite{KulkarniRotating2020}.

We now match the calculation of the density in the two regimes $\Omega = O(1)$ and $1-\Omega = O(1/N)$. 
We show that the corresponding formula for the density match when $\rho(r) = O(\mu)$. 
In the second regime, as we found above $\rho(r)= \frac{k}{\pi}$, where 
$k$ is obtained by inverting Eq.~\eqref{eq:ellk}, which 
gives $k \simeq \frac{\mu-\left(1-\Omega\right)\ell_{k}}{2}$. In the large $k$ limit
one can approximate $\ell_k \sim r^2$ which leads to 
\be
\rho\left(r\right)\simeq\frac{\mu-\left(1-\Omega\right)r^{2}}{2\pi} \; .
\ee
It is easy to check that this is in agreement with Eq.~\eqref{eq:DensityOmegaO1} in the limit $1-\Omega\ll1$, 
so the two regimes indeed match smoothly.

We now turn to the calculation of the variance of ${\cal N}_{R}$, following similar steps to those we used when we calculated the density in Eqs. 
\eqref{density1}-\eqref{densfinal}. Using \eqref{sum2} and the HO approximation \eqref{Vell2} 
\bea
\label{eq:VarNRIntegral}
&& {\rm Var}\,{\cal N}_{R}\simeq\sum_{k=1}^{n+1}\int_{\ell_{k}}^{\ell_{k-1}} \! d\ell\,{\rm Var}\left.{\cal N}_{\left[-\sqrt{\ell},R-\sqrt{\ell}\right]}\right|_{2 u^{2},k}\\
&& \simeq2R \, \theta\left(\sqrt{\ell_{k}}<R<\sqrt{\ell_{k-1}}\right)  \! \int_{-\infty}^{+\infty} \!\! da \, {\rm Var} \left.{\cal N}_{(-\infty,a]}\right|_{2u^{2},k}   \,, \nn
\eea
where this time we used the change of variable $a=R - \sqrt{\ell}=O(1)$, with $\ell =O(N)$ at large $N$.
We thus see that the variance
is proportional to $R$ (the same applies to higher cumulants), and we will now calculate the
prefactor.
Using \cite{footnote:rescaling} 
\be
\label{eq:Rescaling}
\left.{\cal N}_{(-\infty,a]}\,\right|_{2u^{2},m_{\ell}}=\left.{\cal N}_{\left(-\infty,\sqrt{2}a\right]}\,\right|_{\frac{1}{2}u^{2},m_{\ell}}
\ee
we thus obtain that the variance is a discontinuous piecewise linear function
\be
\label{eq:VarNR}
{\rm Var}\,  {\cal N}_R \simeq \sqrt{2} \, C_k R \quad , \quad \sqrt{\ell_k} < R < \sqrt{\ell_{k-1}}
\ee
where we recall that the $\ell_k$'s are given in \eqref{eq:ellk}}. 
Here the
\be \label{Ck}
C_{k}=\int_{-\infty}^{+\infty}da\,{\rm Var}\left.{\cal N}_{(-\infty,a]}\,\right|_{\frac{1}{2} u^{2},k}
\ee 
are numbers that we have calculated in Appendix \ref{app:Ck}. They are given by the explicit formula for integer $k \geq 1$
\be \label{Ck_exp} 
C_{k+1}=\frac{\Gamma(k+\frac{1}{2})^{2}}{\sqrt{2\pi}\pi k!^{2}}\,_{3}F_{2}\left(\frac{3}{2},-k,-k;\frac{1}{2}-k,\frac{1}{2}-k;1\right)
\ee 
where $_{3}F_{2}\left( \cdots \right)$ denotes the hypergeometric function, with $C_k = \frac{1}{\sqrt{2 \pi}} \{1,\frac{7}{4},\frac{153}{64},\dots\}$.
In Fig.~\ref{FigVarNR}, numerical simulations of ${\rm Var}\,  {\cal N}_R$ are compared with the prediction of \eqref{eq:VarNR}. The agreement is good except in the boundary layers $R \simeq \sqrt{\ell_k}$.
%
%
From the above formula we can extract the asymptotics of $C_k$ for large $k$, see Appendix \ref{app:Ck},
\be
\label{eq:BnAsymptotic}
C_{k}\simeq\frac{\sqrt{2}}{\pi^{2}}\sqrt{k}\left(\log k+6\log2+\gamma_{E}-2\right),\quad k\gg1\,.
\ee

\begin{figure}[ht]
\centering
\includegraphics[angle=0,width=0.48\textwidth]{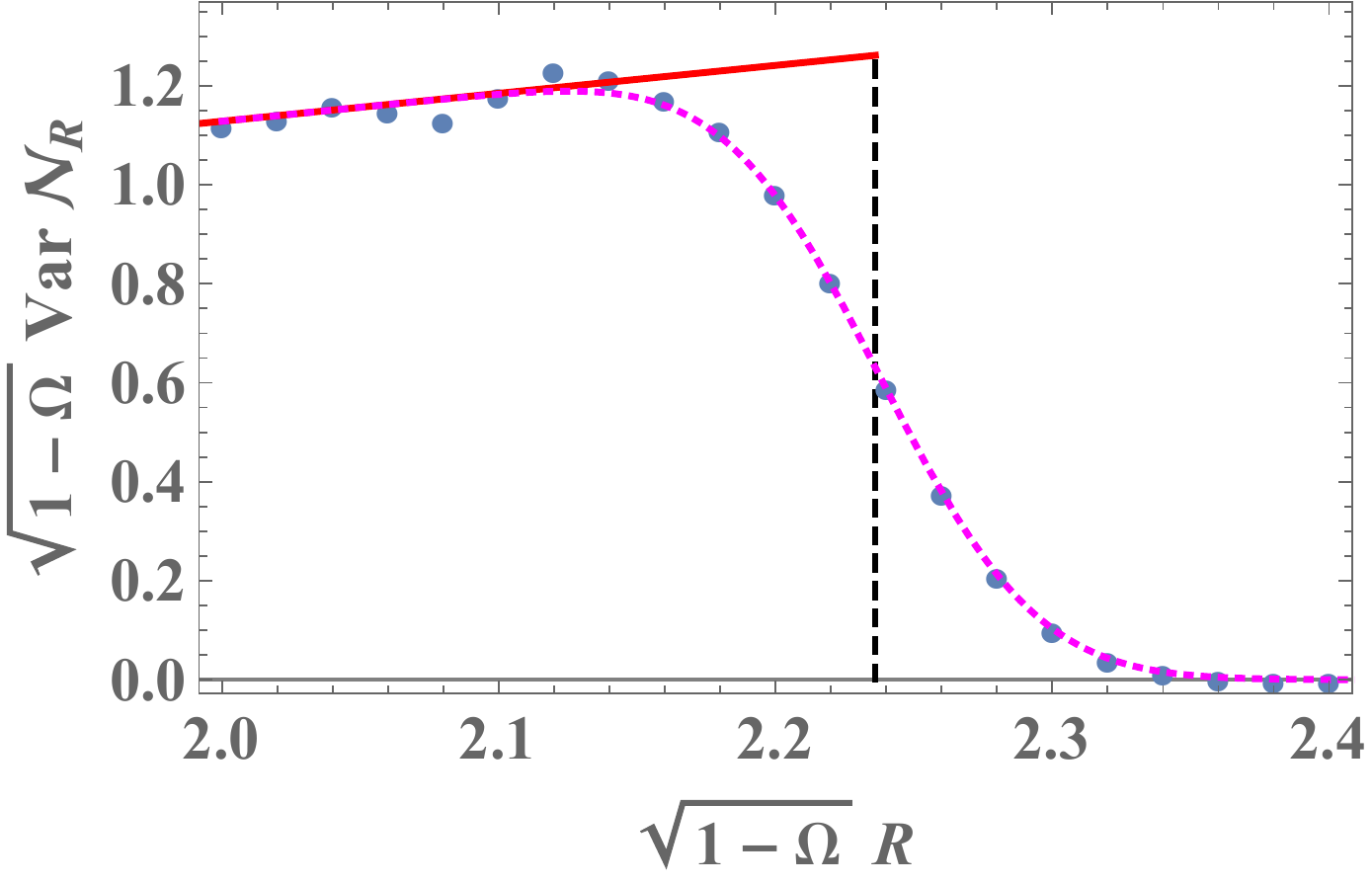}
\caption{
A zoom in on the vicinity of $R=R_e$ of the data plotted Fig.~\ref{FigVarNR} (b) for $\Omega = 0.99$ (blue circles), which is seen to be in excellent agreement with the magenta dotted line corresponding to the edge prediction \eqref{f0EdgeVar}.
}
\label{FigVarNRWithf0Edge}
\end{figure}

In Appendix \ref{sec:stepStructureVar} we obtain the structure of the boundary layer describing the step at $R \simeq \sqrt{\ell_k}$. It takes the form
\be
\label{eq:VarNRStep}
{\rm Var}\,{\cal N}_{R}\simeq\sqrt{2}R\left[C_{k}+f_{k}^{\text{edge},\text{Var}}\left(\sqrt{2}\left(R-\sqrt{\ell_{k}}\right)\right)\right]
\ee
where the scaling function $f_{k}^{\text{edge},\text{Var}}(s)$ depends on $k$, and its
exact expression for any $k \geq 0$ is given in \eqref{fkEdgeVar}.
It is such that $f_{k}^{\text{edge},\text{Var}}\left(s\to -\infty\right)=C_{k+1}-C_k$, and 
$f_{k}^{\text{edge},\text{Var}}\left(s\to +\infty\right)=0$, so as to match smoothly with the 
result within a plateau. In particular, this implies that the width of each boundary layer is of order unity.
The true edge of the Fermi gas (beyond which the total density vanishes) is at $R\simeq  \sqrt{\ell_0}$ and corresponds to $k=0$.
In that case
Eq.~\eqref{eq:VarNRStep} is also valid near this edge, with $C_0=0$, and with
\bea
\label{f0EdgeVar}
&& f_{0}^{\text{edge},\text{Var}}\left(s\right)= \\
&& \frac{1}{4}\left[2 \sqrt{\frac{1}{\pi}}e^{-s^{2}}\text{erf}\left(s\right)+ s \, \text{erf}\left(s\right)^{2}+ \sqrt{2} \frac{\text{erfc}(\sqrt{2}s )}{\sqrt{\pi}}- s \right] \, . \nn
\eea
The prediction \eqref{f0EdgeVar} is plotted opposite numerical data from simulations in Fig.~\ref{FigVarNRWithf0Edge}, showing excellent agreement.
In the LLL limit $\Omega \to 1$, we find that our result is in agreement with \cite{LMG19}, the connection between the scaling functions being $f_{0}^{\text{edge},\text{Var}}\left(s\right)=\mathcal{K}_{2}^{\text{e}}\left(s\right)$ as defined in \cite{LMG19} (see also \cite{CharlierGin}), 
note however \cite{footnote:typoLMG19}. Here we obtain the shape of the steps for all plateaus $k \geq 0$, using 
a rather different method.

Although we have not attempted to compute exactly the higher cumulants of ${\cal N}_{R}$, we can give some general properties. 
The formula \eqref{eq:VarNRIntegral} extends to higher cumulants, which implies that all cumulants
have the form of a piecewise linear function of $R$, namely
\be
\label{eq:CumNRIntegral}
\left\langle {\cal N}_{R}^{2m}\right\rangle ^{c}\simeq\sqrt{2}\ \,C_{k,2m}\,R\,\theta\left(\sqrt{\ell_{k}}<R<\sqrt{\ell_{k-1}}\right)
\ee
where the coefficients $C_{k,2m}$ can be obtained from the number cumulants for the standard $1d$ harmonic oscillator
\be \label{Ck2m}
C_{k,2m}=\int_{-\infty}^{+\infty}da\,\left.\left\langle {\cal N}_{(-\infty,a]}^{2m}\right\rangle ^{c}\right|_{\frac{1}{2}u^{2},k}\;,
\ee 
with $C_{k,2} = C_k$ given in Eqs. (\ref{Ck}) and (\ref{Ck_exp}). They are
discussed in Appendix \ref{sec:higher}. Although we have not studied these coefficients in detail, 
we show, in section \ref{sec:matching} that their large $k$ asymptotics 
match the $1-\Omega=O(1)$ result \eqref{cumulants} in the limit $1-\Omega \ll 1$, as we also show explicitly in the case of the variance.

Finally, from the above predicted form of the cumulants, together with the relation \eqref{SqSumCumulants}, we conclude that the entanglement entropy should also
be a piecewise linear function of $R$ in this regime. More precisely 
\be \label{Sq}
S_{q}\left(D_{R}\right) \simeq  \sqrt{2}\ \, \gamma_{q,k} \, R\,   \theta\left(\sqrt{\ell_{k}}<R<\sqrt{\ell_{k-1}}\right) 
\ee 
where the prefactors $\gamma_{q,k}$'s are related to the bipartite entanglement entropy 
$S_q\left((-\infty,a],\frac{1}{2} u^{2},k\right)$
of the interval $(-\infty,a]$ for the standard $1d$ harmonic oscillator with Hamiltonian $- \frac{1}{2} \frac{\partial^2}{\partial u^2} + \frac{1}{2} u^2$
in its ground state with $k$ fermions, namely
\be \label{gammaq}
\gamma_{q,k} = \int_{-\infty}^{+\infty}da\, S_q\left((-\infty,a],\frac{1}{2} u^{2},k\right) \;.
\ee 
In the special case $k=1$, the entanglement entropy of a single fermion $S_q\left((-\infty,a],\frac{1}{2} u^{2},k\right)$ can be straightforwardly computed from the overlap matrix $A$ defined in (\ref{eq:defA}), using that $S_q= \frac{1}{1-q} {\rm Tr} \log(A^q + (1-A)^q)$, and yielding 
\bea 
&&S_q\left((-\infty,a],\frac{1}{2} u^{2},k=1\right) \nn\\
&&\qquad = \frac{1}{1-q} \log \left( \frac{1}{2^q} {\rm erfc}(a)^q + \frac{1}{2^q} {\rm erfc}(-a)^q \right) \;,
\eea
which, together with Eqs. (\ref{Sq}) and (\ref{gammaq}), coincides with the result obtained in \cite{LMG19}.

\section{Matching the two regimes in the bulk}\label{sec:matching}

It is now interesting to match (i) the result for the variance in \eqref{eq:VarNR}, \eqref{Ck},
which is valid in the vicinity of the LLL,
i.e. for $1-\Omega \simeq c/N$ with $c=O(1)$ and $\mu=O(1)$, with (ii) the result in \eqref{eq:VarNROmegaO1} which is valid in the limit $1-\Omega\ll1$ and
in the bulk of the Fermi gas. Let us consider now $1-\Omega \simeq c/N$ with $c \to +\infty$. Comparing
Eqs. (\ref{mu}) and (\ref{mu2}) we see that it also corresponds to large $\mu$, with $c \sim \mu^2/4$. We recall that
$k$ varies from $k=0$ (at the edge of the Fermi gas) to $k=k_{\rm max}=n+1 \simeq \mu/2$ near the center.
Since we are interested in the bulk, $k=O(\mu)$ is large and we can use the estimate $k= \frac{\mu - (1- \Omega) \ell_k}{2}$.
Using \eqref{eq:VarNR} and the asymptotic estimate \eqref{eq:BnAsymptotic} we obtain
\bea
\label{eq:VarNRMatching}
&& \!\!\!\! {\rm Var}\,{\cal N}_{R} \simeq \mu\tilde{R}\frac{2}{\pi^{2}}\sqrt{1-2\left(1-\Omega\right)\tilde{R}^{2}} \times \nn\\
&& \!\!\!\! \left[\log\mu+\log\left(1-\left(1-\Omega\right)2\tilde{R}^{2}\right)+5\log2+\gamma_{E}-2\right]  .
\eea
where $\tilde R=R/\sqrt{2 \mu}$ is fixed, and we have used $\ell_k \sim R^2$. 
On the other hand, one can consider Eq.~\eqref{eq:VarNROmegaO1} when $1-\Omega\ll1$ and $R/R_e$ fixed, i.e., 
$\tilde{R}\sim R_{e}/\sqrt{2\mu}=1/\sqrt{1-\Omega^{2}}\gg1$. As we show in the Appendix \eqref{eq:VarNRMatching2} 
in this limit the result matches precisely with Eq. \eqref{eq:VarNRMatching}.

Similarly, we can match the behavior of the higher cumulants in the two regimes. Although we did not calculate the constants $C_{k,2m}$ for $m\geq 2$, from Eq.~\eqref{Ck2m} explicitly, we now find their large-$k$ behavior using a heuristic argument. At $k\gg1$, the integrand in \eqref{Ck2m} takes a constant value in the bulk of the effective 1d fermi gas and vanishes outside the bulk \cite{UsCounting2020} for $m\geq 2$:
\be
\left.\left\langle {\cal N}_{(-\infty,a]}^{2m}\right\rangle ^{c}\right|_{\frac{1}{2}u^{2},k}\simeq\theta\left(\left|a\right|<\sqrt{2k}\right)\frac{\kappa_{2m}}{2} \, .
\ee
where $\kappa_{2m}$ is defined in Eq.~\eqref{kappDef}, and the factor $1/2$ comes from the fact that it is a semi-infinite interval. Thus, Eq.~\eqref{Ck2m} simply yields $C_{k,2m}\simeq\sqrt{2k}\,\kappa_{2m}$.
As in the case of the variance, we now use $k=\frac{\mu-(1-\Omega)\ell_{k}}{2}$ which, together with \eqref{eq:CumNRIntegral}, yields for $m\geq 2$
\be
\label{eq:CumMatching}
\left\langle {\cal N}_{R}^{2m}\right\rangle ^{c}\simeq R\sqrt{2\mu-2(1-\Omega)R^{2}}\,\kappa_{2m} \, .
\ee
On the other hand, one can now take the limit $1 - \Omega \ll 1$ in \eqref{cumulants}, and it is straightforward to see that the result is in perfect agreement with \eqref{eq:CumMatching}.
 Note that this heuristic argument can also be performed for the variance and reproduces the above results
\cite{CkAsymptotic}.

\section{Extension to 3d geometry}\label{sec:3d}

We now consider the Hamiltonian \eqref{eq:H3d}. In cylindrical coordinates $(r,\phi,z)$, the eigenfunctions of $H$ are given by $\Psi\left(r,\phi,z\right)=\psi\left(r,\phi\right)\xi\left(z\right)$ where $\psi\left(r,\phi\right)$ are the eigenfunctions of the 2d Hamiltonian \eqref{eq:H2d} and $\xi\left(z\right)$ are eigenfunctions of the $z$-component Hamiltonian, $H_{z}=\frac{p_{z}^{2}}{2}+U\left(z\right)$.
or a system of $N$ noninteracting fermions in the ground state, one simply fills up the $N$ lowest single particle eigenstates.

Let ${\cal N}_{R}$ now be the number of particles in the cylinder $\left\{ \left(r,\phi,z\right);\;r<R\right\}$ of radius $R$ around the origin, fully extended in the $z$ direction.
Then one finds that the statistics of ${\cal N}_{R}$ decompose in the different sectors of $H_z$ in a manner similar to the angular decomposition described above in \eqref{1}.
In particular, the radial density, defined by
\be
\rho_{\text{rad}}\left(r\right)=\int_{-\infty}^{\infty}\rho_{3d}\left(r,z\right)dz
\ee 
where $\rho_{3d}\left(r,z\right)$ is the density of the 3d gas, is given by
\be
\rho_{\text{rad}}\left(r\right)=\sum_{\epsilon_{z}}\rho_{2d}^{\left(\mu-\epsilon_{z}\right)}\left(r\right)
\ee
where $\rho_{2d}^{\left(\mu-\epsilon_{z}\right)}\left(r\right)$ is the density of an effective $2d$ fermi gas with single-particle Hamiltonian \eqref{eq:H2d} and fermi energy $\mu-\epsilon_{z}$.
Similarly, the variance of ${\cal N}_{R}$ is given by
\be
\label{VarNR3d}
{\rm Var}\,{\cal N}_{R}=\sum_{\epsilon_{z}}{\rm Var}\,{\cal N}_{R,\left(2d\right)}^{\left(\mu-\epsilon_{z}\right)}
\ee
where ${\cal N}_{R,\left(2d\right)}^{\left(\mu-\epsilon_{z}\right)}$ is the number of particles in a circle of radius $R$ for a $2d$ fermi gas with single-particle Hamiltonian \eqref{eq:H2d} and fermi energy $\mu-\epsilon_{z}$.
These decompositions into the different sectors of $H_z$ extend to all higher cumulants as well, and therefore also to the entanglement entropy.

To illustrate these general results, let us consider an anisotropic harmonic oscillator described by $V(r) = r^2 / 2$ and $U(z) = \nu^2 z^2 /2$. Then Eq.~\eqref{VarNR3d} becomes
\be
{\rm Var}\,{\cal N}_{R}=\sum_{i_{z}=0}^{i_{\max}}{\rm Var}\,{\cal N}_{R,\left(2d\right)}^{\mu-\left(i_{z}+\frac{1}{2}\right)\nu}
\ee
with $i_{\max}=\text{Int}\left(\frac{\mu-1}{\nu}-\frac{1}{2}\right)$, and at $N\gg1$, ${\rm Var}\,{\cal N}_{R,\left(2d\right)}^{\mu-\left(i_{z}+\frac{1}{2}\right)\nu}$ is given by \eqref{eq:VarNROmegaO1} and/or by \eqref{eq:VarNR} [replacing $\mu \to \mu-\left(i_{z}+\frac{1}{2}\right)\nu$], depending on the scaling of $\Omega$ with $N$.

\section{Conclusion}

In summary, we studied the ground state of $N \gg 1$ noninteracting 
fermions in a two-dimensional harmonic trap of frequency $\omega$, in a reference frame 
which is rotating at angular frequency $\Omega<\omega$ around the origin. 
We calculated the FCS of the number of fermions ${\cal N}_R$ inside a disk of radius $R$ centered at the origin, 
within the bulk of the Fermi gas, i.e. for
$R/R_e = O(1)$ where $R_e$ is the edge radius. We found rich and interesting behaviours in the two different scaling regimes: $\Omega / \omega = O(1)$ and 
$1 - \Omega / \omega = O(1/N)$.
In the latter regime we find that ${\rm Var}\,{\cal N}_{R}$ is given by a discontinuous piecewise linear function of $R$,
of the form $\sim (R/R_e) \sqrt{N}$. This behavior has the same origin as the ``wedding cake'' structure found for the density in \cite{KulkarniRotating2020}, namely the successive filling of Landau levels. We have shown how, upon decreasing 
$\Omega / \omega$, this piecewise linear function becomes a continuous function of $R$ with ${\rm Var}\,{\cal N}_{R}\sim \sqrt{N} \log N$
whose precise form we obtained analytically in the whole regime $\Omega/\omega =O(1)$. This prediction has also been tested numerically, see Fig.~\ref{FigVarNR}.
We also studied the properties of the gas near the edge, and checked our results numerically, see Fig.~\ref{FigVarNRWithf0Edge}.
Similar results hold for higher cumulants, namely they are piecewise linear functions of $R$ in the regime $1 - \Omega / \omega = O(1/N)$,
while in the regime $\Omega/\omega =O(1)$ they are all proportional to a smooth function of $R$ obtained here. 
As an application, we also calculated the bipartite entanglement entropy of a disk of radius $R$ centered at the origin with the rest of the system.
It exhibits a similar change of behavior as $\Omega$ is decreased, and in the regime $1 - \Omega / \omega = O(1/N)$ we predict that it is also a piecewise linear
function of $R$. Finally, we showed how the FCS can be studied in a three-dimensional extension of the model, where an additional confining potential is applied in the $z$ direction.

Another outcome of our results is that we obtain, by the same calculation, the FCS in $p$ space. Indeed the 2D Hamiltonian \eqref{Ham2d}
is invariant by the transformation $H({\bf p},{\bf r}) = H({\bf r},-{\bf p})$ which conserves the commutator. Hence the full kernel (not studied here) 
is the same in real space and in momentum space (in the LLL it is related to the kernel for the Ginibre ensemble). 
This implies that the momentum density has the same structure (at large $N$ uniform, with plateaus) as the one in real space,
and that the cumulants are identical. This remark is of interest for time of flight experiments which measure the
density in momentum space \cite{Pagano14}.

There are several avenues left for future research. A natural next step would be to calculate the coefficients $C_{k,2m}$ in \eqref{Ck2m} associated 
to the piecewise linear dependence of the higher cumulants in the regime $1 - \Omega / \omega = O(1/N)$, 
and therefore, by virtue of \eqref{SqSumCumulants}, also obtain the exact coefficients $\gamma_{q,k}$ for the entanglement entropy,
thus extending the results for $\gamma_{q,1}$ for the lowest Landau level in \cite{LMG19}. 
It would also be interesting to study the full distribution of ${\cal N}_R$, 
and in particular its large-deviation form, 
as was done in \cite{LaCroixGinibrePRE} for the Ginibre ensemble (corresponding to $\Omega \to \omega$ in our system),
 see also \cite{Allez14,CharlierAnnuli} in the context of Ginibre matrices.

Our results can be extended to more general trapping potentials $V(r)$. For $\Omega <\omega$
it is possible to apply the method developed in our previous work at $\Omega=0$
\cite{UsCounting2020} which was able to treat an arbitrary external potential.
To explore further the regime $1- \frac{\Omega}{\omega}=O(1)$ one could
consider a class of external potentials $V(r)$ which almost cancel the centrifugal potential, 
so that only a few Landau levels are occupied. One example was provided in \cite{KulkarniRotating2020} but there is a
much larger class of potentials with similar features. In that case the FCS can still be obtained
from the approximation by the $1d$ harmonic oscillator used in the present work. As a consequence our results should
exhibit universal features for this larger class of potentials. 
Another interesting example is a potential $V(r)$ that contains anharmonic terms, in addition to a harmonic term:
one could consider the case $\Omega/\omega > 1$ where the effective radial potential can become a double well. In such situations it is known
that the FCS is anomalous near the transition to the double well, see \cite{smi20} for an example in $1d$. 

Additional important extensions of our results would be (i) to systems of interacting fermions, as was recently done for some particular 1d systems \cite{InteractingFCS}, and (ii) to nonzero temperature, where one could use the framework of \cite{DeanPLDReview}. These extensions could be very useful towards a theory that describes more accurately systems studied in experiments.

\bigskip

{\it Acknowledgments:} 
We thank M. Kulkarni and B. Lacroix-A-Chez-Toine for previous collaborations on related topics. NRS acknowledges support from the Yad Hanadiv fund (Rothschild fellowship). 
This research was supported by ANR grant ANR-17-CE30-0027-01 RaMaTraF.

\begin{widetext} 

\appendix

\section{Semi-classical density}
\label{sec:semiclassical}

From general results, the Wigner function in the bulk and for large $N$ is given by
the semi-classical formula \cite{Wigner18} 
\be
W({\bf r},{\bf p})\simeq\frac{1}{(2\pi)^{2}}\theta\left(\mu-\frac{1}{2}\left({\bf p}-\Omega\,\hat{{\bf z}}\wedge{\bf r}\right)^{2}-V(r)+\frac{\Omega^{2}}{2}r^{2}\right)\;,
\ee
where $\theta(z)$ is the Heaviside step function and $\hat {\bf z}$ is the unit vector along the $z$ axis. The mean density can be obtained from this formula using $\rho({\bf r})= \int d^{2}{\bf p}\, W({\bf r},{\bf p})$.

In the case of the harmonic potential $V(r) = \frac{1}{2} r^2$, this yields
\bea \label{rhoreal} 
&& \rho({\bf r})\simeq \frac{1}{(2\pi)^{2}}\int d^{2}{\bf p}\,\theta\left(\mu-\frac{1}{2}({\bf p}-\Omega\,\hat{{\bf z}}\wedge{\bf r})^{2}-\frac{(1-\Omega^{2})}{2}r^{2}\right)=\frac{1}{(2\pi)^{2}}\int d^{2}{\bf p}\,\theta\left(\mu-\frac{1}{2}{\bf p}^{2}-\frac{(1-\Omega^{2})}{2}r^{2}\right) = \frac{k_F(r)^2}{4 \pi} \, , \nn\\\\
&& k_F(r) = \sqrt{2 \mu - (1-\Omega^2) r^2 }
\eea
obtained upon shifting the ${\bf p}$ variable. We have introduced the local 2D Fermi momentum $k_F(r)$. 
These results are given in the main text in Eq.~\eqref{eq:DensityOmegaO1}.
The total number of particles is
\be
N=\int\rho\left({\bf r}\right)d^{2}{\bf r}\simeq\frac{1}{4\pi}\int d^{2}{\bf r}\left[2\mu-\left(1-\Omega^{2}\right)r^{2}\right]=\frac{\mu^{2}}{2\left(1-\Omega^{2}\right)}
\ee 
in agreement with \eqref{rel_Nmu} which was obtained in the main text using a different method.

Note that the mean density in momentum space, $\hat \rho({\bf p})$, has a very similar form. Indeed, by rearranging the terms we obtain 
\be \label{rhomom} 
\hat \rho({\bf p})\simeq \frac{1}{(2\pi)^{2}}\int d^{2}{\bf r}\,\theta\left(\mu
-\frac{1}{2}(x - \Omega p_y)^2 
-\frac{1}{2}(y + \Omega p_x)^2 
-\frac{(1-\Omega^{2})}{2}r^{2}\right) = \frac{2 \mu - (1- \Omega^2) p^2}{4 \pi} \;.
\ee 

{\bf Remark}. Note that in the quite different limit of the LLL, i.e. $\Omega \to 1$, the kernel at large $N$ and in the bulk is 
given by the kernel of the Ginibre ensemble of random matrix theory, namely
$K(z,z') = \frac{1}{\pi} e^{- \frac{1}{2} |z|^2 - \frac{1}{2} |z'|^2 + z (z')^* }$
with $z=x+ i y$. One can check, by simple Fourier transform, that in $p$ space it retains a similar form than in real space. 
This, as well as the fact that \eqref{rhomom} and \eqref{rhoreal} are very similar, are two manifestation of
the symmetry $H({\bf p},{\bf r}) = H({\bf r},-{\bf p})$ of the Hamiltonian, as discussed in the conclusion of this paper.
This shows that correlations in real and momentum space have identical forms (which is an exact property valid for any $N$).

\section{FCS for the equivalent harmonic oscillator problem} 
\label{app:Ck}

In this section we compute the coefficients $C_k$ defined in the text. 
Consider the eigenfunction $\varphi_j$'s of the harmonic oscillator $H = \frac{p^2}{2} + \frac{x^2}{2}$ 
\bea \label{def_hermite_function}
\varphi_j(x) = \frac{1}{\sqrt{\sqrt{\pi} \, 2^j j!}} H_j(x) e^{-x^2/2}  \quad , \quad j=0,1,\dots
\eea 
One defines the overlap matrix and its complementary
\be \label{eq:defA} 
A_{ij}(a)= \int_{-\infty}^a dx \, \varphi_i(x) \varphi_j(x) \quad , \quad \delta_{ij}- A_{ij}(a)=  \int_{a}^{+\infty} dx \, \varphi_i(x) \varphi_j(x) \;.
\ee 

We aim to calculate the integrated variance when the oscillator is occupied up to level $j=n$,
which corresponds to $m_\ell=n+1$ fermions and to the coefficient $C_{k=n+1}$ defined in the text. Using the standard
formula which relates the variance to the overlap matrix (see e.g. \cite{CMV2012}) we have 
\be \label{defC} 
C_{n+1}= B_n = \int_{-\infty}^\infty da \sum_{i,j=0}^n A_{ij}(a) (\delta_{ij}- A_{ij}(a)) 
\ee 
where we have defined $B_n=C_{n+1}$ for convenience. 
We first calculate the generating function 
\bea
\tilde B(z,t) = \int_{-\infty}^\infty da \sum_{i , j \geq 0} z^i t^j A_{ij}(a) (\delta_{ij}- A_{ij}(a)) = 
\sum_{i , j \geq 0} z^i t^j  \int_{-\infty}^\infty da \int_{-\infty}^a dx \, \varphi_i(x) \varphi_j(x) 
\int_a^{+\infty} dy \, \varphi_i(y) \varphi_j(y)  \; .
\eea 
Using the Mehler formula 
\bea \label{Mehler0}
\sum_{j=0}^\infty \frac{z^j}{2^j j!} H_j(x) H_j(y) = G(x,y,z) =
\frac{1}{\sqrt{1-z^2}} e^{\frac{2 x y z-(x^2+y^2)z^2}{1-z^2}} \;,
\eea
the generating function can be written as
\be
\tilde B(z,t) = \frac{1}{\pi} \int_{-\infty}^\infty da \int_{-\infty}^a dx 
\int_a^{+\infty} dy \, G(x,y,z) G(x,y,t) e^{- x^2 - y^2} \;.
\ee 

To perform this integral we shift the arguments $x,y$ by defining $x=a - X$ and $y= a + Y$,
where now $X,Y$ are both
integrated on $[0,+\infty[$. The integral over $a$ is an elementary Gaussian integral and
leads to a simple Gaussian integrand depending only on $X+Y$. Its integration is elementary and leads to
\be
\tilde B(z,t) = \frac{1}{\sqrt{2 \pi}} \frac{\sqrt{1-t} \sqrt{1-z}}{(1-t z)^{3/2}}
\ee

From this result we obtain the generating function for the $B_k$ as follows. Define
\be
B(z,t) = \sum_{n_1,n_2 \geq 0} z^{n_1} t^{n_2} \sum_{i=0}^{n_1}
\sum_{j=0}^{n_2}     \int_{-\infty}^\infty da A_{ij}(a) (\delta_{ij}- A_{ij}(a)) 
\ee 
which is such that the $B_n$'s are the diagonal coefficients for $n_1=n_2$
\be
B_n = B(z,t) |_{z^n t^n}      \quad , \quad [ B(z,t) ]_{\rm diag} = \sum_{n \geq 0} (z t)^n B_n
\ee 
On the other hand one has simply
\be \label{Bapp}
B(z,t) = \frac{1}{1-z} \frac{1}{1-t} \tilde B(z,t) = \frac{1}{\sqrt{2 \pi}} \frac{1}{\sqrt{1-t} \sqrt{1-z} (1-t z)^{3/2}} \;.
\ee 

The diagonal part of $B(z,t)$ can be extracted as follows.
Using that $(1+x)^a = \sum_{n \geq 0} {\rm Bin}(a,n) x^n$ where here we denote ${\rm Bin}(a,n)=\frac{\Gamma(a+1)}{\Gamma(n+1) \Gamma(a-n+1)}$ the binomial
coefficient, one has
\be
\left[(1-z)^{-1/2}(1-t)^{-1/2}\right]_{{\rm diag}}=\sum_{n\geq0}B\left(-\frac{1}{2},n\right)^{2}(zt)^{n}=\,_{2}F_{1}\left(\frac{1}{2},\frac{1}{2},1,zt\right)=\frac{2K(\sqrt{zt})}{\pi}
\ee
where $K\left(x\right)=\frac{\pi}{2}\sum_{n=0}^{\infty}\left[\frac{\left(2n-1\right)!!}{\left(2n\right)!!}\right]^{2}x^{2n}$ is the elliptic integral of the first kind. 
This leads to (with $*$ denoting the convolution) 
\be
B_{n}=\frac{(-1)^{n}}{\sqrt{2\pi}} B\left(-\frac{3}{2},n\right)*B\left(-\frac{1}{2},n\right)^{2} =\frac{1}{\sqrt{2\pi}}\sum_{k=0}^{n}(-1)^{n-k}B\left(-\frac{3}{2},n-k\right)B\left(-\frac{1}{2},k\right)^{2}
\ee 
This formula gives, for $n=0,1,\dots$
\be
B_n = {\frac{1}{\sqrt{2 \pi}}} \left\{1,\frac{7}{4},\frac{153}{64},\frac{759}{256},\frac
   {57225}{16384},\frac{261207}{65536},\frac{4667201}{1
   048576},\frac{20525111}{4194304},\frac{5709117897}{1
   073741824},\frac{24584746575}{4294967296},\frac{4203
   44955129}{68719476736},\dots\right\}
\ee 
The corresponding numerical
values are
\be
B_n = \frac{1}{\sqrt{2\pi}}\{1,1.75,2.39063,2.96484,3.49274,3.9857,4.45099,4.89357,5
   .31703,5.72408,6.11682\}
\ee 
Another way to present the result is in the form of the generating function 
\be \label{GF_B}
B(z)=\sum_{n \geq 0} z^n B_n = {\frac{1}{\sqrt{2 \pi}}}\, _2F_1 \left(\frac{1}{2}, \frac{1}{2}, 1,z \right) (1-z)^{-3/2} 
\ee 

To extract a more compact formula for the $B_n$ and the $C_k=B_{k-1}$ given in the text in \eqref{Ck}, we will
use a slightly different method. Starting from \eqref{Bapp} we use an integral represention
\bea
&& B(z,t) = \frac{1}{\sqrt{2 \pi}} \frac{1}{\sqrt{1-t} \sqrt{1-z} (1-t z)^{3/2}} = \frac{1}{\sqrt{2 \pi}} 
\frac{2}{\pi^{3/2}} \int_0^{+\infty} dx_1 dx_2 dx_3 x_1^{-1/2} x_2^{-1/2} x_3^{1/2} 
e^{ - x_1 (1-t) - x_2 (1-z) - x_3 (1- zt)} \nonumber \\
&&=  \frac{1}{\sqrt{2 \pi}} 
\frac{2}{\pi^{3/2}} 
\sum_{n_1,n_2,n_3 \geq 0}
\frac{\Gamma(n_1+\frac{1}{2}) \Gamma(n_2+\frac{1}{2}) \Gamma(n_3+\frac{3}{2}) }{n_1! n_2! n_3!} 
z^{n_1+n_3} t^{n_2+n_3} \\
&& = \frac{1}{\sqrt{2 \pi}} 
\frac{2}{\pi^{3/2}} 
\sum_{n_1,n_2 \geq 0} z^{n_1} t^{n_2}  \sum_{n_3=0}^{\min(n_1,n_2)} 
\frac{\Gamma(n_1-n_3+\frac{1}{2}) \Gamma(n_2-n_3+\frac{1}{2}) \Gamma(n_3+\frac{3}{2}) }{(n_1-n_3)! (n_2-n_3)! n_3!} 
\eea 
where in the second line we have expanded the exponentials in powers of $z$ and $t$ and integrated over the auxiliary variables $x_i$
and in the third line we have rearranged the terms. The diagonal term $n_1=n_2$ in the double series provides the expression
for $B_n$
\bea
B_n =  \frac{1}{\sqrt{2 \pi}} 
\frac{2}{\pi^{3/2}}   \sum_{n_3=0}^{n} 
\frac{\Gamma(n-n_3+\frac{1}{2})^2 \Gamma(n_3+\frac{3}{2}) }{[(n-n_3)!]^2 n_3!} 
\eea 
Using mathematica 
this sum can be performed and leads to 
\be
C_{n+1}= B_n = \frac{1}{\sqrt{2 \pi}} \frac{\Gamma \left(n+ \frac{1}{2}\right)^2 \,
   _3F_2\left(\frac{3}{2},-n,-n;\frac{1}{2}-n,\frac{1}{
   2}-n;1\right)}{\pi n!^2}
\ee
which leads to the result \eqref{Ck} given in the text for $C_k$. 
\\

Let us now study the asymptotic behavior of $C_k$ for large $k$. It can be obtained from the behavior of the generating function $B(z)$ in Eq. (\ref{GF_B}) for $z \to 1$. In that
limit one finds
\be
B(z)\simeq\frac{1}{\sqrt{2}\,\pi^{3/2}}\frac{1}{(1-z)^{3/2}}\left[-\log(1-z)+4\log2\right] \;.
\ee 
By setting $z=e^{-s}$ one obtains
\be \label{Bn_large}
B_n \simeq \frac{\sqrt{2}}{ \pi^2} \sqrt{n} ( \log n + b ) \quad , \quad b = 6 \log 2 + \gamma -2 \;,
\ee
which gives the result \eqref{eq:BnAsymptotic} mentioned in the text for $C_k$. 
To obtain this result \eqref{Bn_large}, we used the relation
\be
\sum_{n}z^{n}\sqrt{n}\left(\log n+c\right)\simeq\int_{0}^{+\infty}dn \, e^{-sn}\sqrt{n}\left(\log n+c\right)=\frac{\sqrt{\pi}}{2s^{3/2}}\left(-\log s-2\log2-\gamma+2+c\right) \;.
\ee

\section{Density and variance near the steps $R \simeq \sqrt{\ell_k}$}

\subsection{The step structure in the density}
\label{sec:stepStructureRho}

Here we calculate the structure of the step, recovering the results of \cite{KulkarniRotating2020} by a different method. In \cite{KulkarniRotating2020}
the result was obtained by first writing the exact formula for the density in the exact eigenfunctions, in terms of the Laguerre polynomials and in a second stage taking the large $\mu$ limit and using standard asymptotic formulas for the Laguerre polynomials in terms of Hermite polynomials. 
Here we use, more directly, that at large $\mu$ the problem can be mapped to a harmonic oscillator. This also provides a test
of the approach in the text. 

Starting from \eqref{eq:DensityOmegaNearOne} in the main text, we zoom in around the step by assuming that $r\simeq\sqrt{\ell_{k}}$ for some $k$. Thus, in the sum in Eq.~\eqref{eq:DensityOmegaNearOne} only the terms with $k$ and $k+1$ contribute,
\be
2\pi r\rho\left(r\right)\simeq\int_{\ell_{k}}^{\ell_{k-1}}d\ell\,\rho_{1d\text{HO},k}\left(r-\sqrt{\ell}\right)+\int_{\ell_{k+1}}^{\ell_{k}}d\ell\,\rho_{1d\text{HO},k+1}\left(r-\sqrt{\ell}\right) \, .
\ee
Adding and subtracting the term $\int_{\ell_{k+1}}^{\ell_{k}}d\ell\,\rho_{1d\text{HO},k}\left(r-\sqrt{\ell}\right)$ to the r.h.s. of the last equation, one obtains $2\pi r\rho\left(r\right) \simeq I_1 + I_2$ where
\be
I_1 = \int_{\ell_{k+1}}^{\ell_{k-1}}d\ell\,\rho_{1d\text{HO},k}\left(r-\sqrt{\ell}\right) \simeq 2kr
\ee
[the last approximate equality is obtained using the same approximation as in Eqs.~\eqref{dens22} and \eqref{densfinal}] and 
\be
\label{eq:I2def}
I_2 = \int_{\ell_{k+1}}^{\ell_{k}}\left[\rho_{1d\text{HO},k+1}\left(r-\sqrt{\ell}\right)-\rho_{1d\text{HO},k}\left(r-\sqrt{\ell}\right)\right]d\ell \, .
\ee
In order to calculate $I_2$, we now (i) change variables $\sqrt{\ell}=r-u$ where we assume $u \ll r$ so $d\ell=2\left(u-r\right)du\simeq-2rdu$ [this is the same change of variables that is described below \eqref{densfinal}] and (ii) recall that
\be
\rho_{1d\text{HO},k}\left(u\right)=\sum_{i=0}^{k-1}\left|\psi_{i}\left(u\right)\right|^{2}
\ee
where $\psi_i(x)$ is the eigenfunction function associated to the $i^{\text{th}}$ energy level of the 1d harmonic oscillator
$H_{\rm 1D\text{HO}}= - \frac{1}{2} \partial_u^2 + 2 u^2$. Putting it all together we find
\be
\label{eq:I2a}
I_{2}\simeq2r\int_{r-\sqrt{\ell_{k}}}^{r-\sqrt{\ell_{k+1}}}\left|\psi_{k}\left(u\right)\right|^{2}du \, .
\ee
Now using 
\be
\psi_{k}\left(u\right)=\left(\frac{\sqrt{2}}{\sqrt{\pi}2^{k}k!}\right)^{1/2}e^{-u^{2}}H_{k}(\sqrt{2} u)
\ee
where $H_{k}(x)$ is the Hermite polynomial of degree $k$, and replacing the upper limit of the integral \eqref{eq:I2a} by infinity (which is a good approximation due to the assumption $r\simeq \sqrt{\ell_k}$) we obtain the density around the step i.e. for $r- \sqrt{\ell_k} = O(1)$
\be
\label{eq:densityStep}
\rho\left(r\right)\simeq\frac{k}{\pi}+f_{k}^{\text{edge}}\left(\sqrt{2} (\sqrt{\ell_{k}}-r) \right), \qquad f_{k}^{\text{edge}}\left(x\right)=\frac{1}{\pi^{3/2}2^{k}k!}\int_{-x}^{\infty}e^{-v^{2}}\left[H_{k}(v)\right]^{2}dv,
\ee
in perfect agreement with \cite{KulkarniRotating2020} (see also \cite{Dunne94,HH2013, FL20}).

\subsection{The step structure in the variance}
\label{sec:stepStructureVar}

We begin from the first line in Eq.~\eqref{eq:VarNRIntegral}, and assume that $R \simeq \sqrt{\ell_k}$ for some $k$. 
Again we zoom in around the step by assuming that only the terms with occupation numbers $k$ and $k+1$ have a non-negligible contribution
to \eqref{eq:VarNRIntegral} 
\be
{\rm Var}\,{\cal N}_{R}\simeq\int_{\ell_{k}}^{\ell_{k-1}}d\ell\,{\rm Var}\,\left.{\cal N}_{\left[-\sqrt{\ell},R-\sqrt{\ell}\right]}\right|_{2u^{2},k}+\int_{\ell_{k+1}}^{\ell_{k}}d\ell\,{\rm Var}\,\left.{\cal N}_{\left[-\sqrt{\ell},R-\sqrt{\ell}\right]}\right|_{2u^{2},k+1} \, .
\ee
Adding and subtracting $\int_{\ell_{k+1}}^{\ell_{k}}d\ell\,{\rm Var}\,\left.{\cal N}_{\left[-\sqrt{\ell},R-\sqrt{\ell}\right]}\right|_{2u^{2},k}$ to the r.h.s. 
and performing the change of variable $a=R - \sqrt{\ell}$, we find
\be \label{aaa} 
{\rm Var}\,{\cal N}_{R} \simeq \sqrt{2}C_{k}R+2R\int_{R-\sqrt{\ell_{k}}}^{\infty}da\,\left({\rm Var}\,\left.{\cal N}_{\left(-\infty,\sqrt{2}a\right]}\right|_{\frac{1}{2}u^{2},k+1}-{\rm Var}\,\left.{\cal N}_{\left(-\infty,\sqrt{2}a\right]}\right|_{\frac{1}{2}u^{2},k}\right)
\ee
where the first term was obtained using the same calculations as those that give Eq.~\eqref{eq:VarNR} in the main text, and we used Eq.~\eqref{eq:Rescaling}.
Note that the upper bound $R - \sqrt{\ell_{k+1}}$ in \eqref{aaa} can be safely set to infinity. 
We now use the known expression for the number variance in determinantal point processes \cite{MehtaBook, Forrester}
\be
{\rm Var}\,\left.{\cal N}_{\left(-\infty,\sqrt{2}a\right]}\right|_{\frac{1}{2}u^{2},k}=\int_{-\infty}^{\sqrt{2}a}dx\int_{\sqrt{2}a}^{\infty}dy\,K_{k}\left(x,y\right)^{2}
\ee
where
\be
K_{k}\left(x,y\right)=\sum_{i=0}^{k-1}\varphi_{i}\left(x\right)\varphi_{i}\left(y\right)
\ee
is the kernel in terms of the eigenfunctions $\varphi_k(x)$ of the harmonic oscillator defined in Eq. (\ref{def_hermite_function}). Thus
\be
K_{k+1}\left(x,y\right)^{2}=\left[K_{k}\left(x,y\right)+\varphi_{k}\left(x\right)\varphi_{k}\left(y\right)\right]^{2}=K_{k}\left(x,y\right)^{2}+2\varphi_{k}\left(x\right)\varphi_{k}\left(y\right)K_{k}\left(x,y\right)+\varphi_{k}^{2}\left(x\right)\varphi_{k}^{2}\left(y\right)
\ee
leading to
\bea
&& {\rm Var}\,\left.{\cal N}_{\left(-\infty,\sqrt{2}a\right]}\right|_{\frac{1}{2}u^{2},k+1}-{\rm Var}\,\left.{\cal N}_{\left(-\infty,\sqrt{2}a\right]}\right|_{\frac{1}{2}u^{2},k} = \int_{-\infty}^{\sqrt{2}a}dx\int_{\sqrt{2}a}^{\infty}dy\,\left[2\varphi_{k}\left(x\right)\varphi_{k}\left(y\right)K_{k}\left(x,y\right)+\varphi_{k}^{2}\left(x\right)\varphi_{k}^{2}\left(y\right)\right] \nn \\
&&\qquad\qquad\qquad\qquad\qquad\qquad\qquad\qquad\qquad =\int_{-\infty}^{\sqrt{2}a}dx\int_{\sqrt{2}a}^{\infty}dy\,2\varphi_{k}\left(x\right)\varphi_{k}\left(y\right)K_{k}\left(x,y\right)+A_k\left(\sqrt{2} a\right)\left[1-A_k\left(\sqrt{2} a\right)\right] \, , \nonumber \\
&& 
\eea
where $A_k\left(v\right)= A_{kk}\left(v\right)=\int_{-\infty}^{v}\varphi_{k}^{2}\left(x\right)dx$ and we used the normalization $\int_{-\infty}^{\infty}\varphi_{k}^{2}\left(x\right)dx=1$. The remaining integral is
\be
\int_{-\infty}^{\sqrt{2}a}dx\int_{\sqrt{2}a}^{\infty}dy\,2\varphi_{k}\left(x\right)\varphi_{k}\left(y\right)K_{k}\left(x,y\right)=2\sum_{i=0}^{k-1}\int_{-\infty}^{\sqrt{2}a}\varphi_{i}\left(x\right)\varphi_{k}\left(x\right)dx\int_{\sqrt{2}a}^{\infty}\varphi_{k}\left(y\right)\varphi_{i}\left(y\right)dy=-2\sum_{i=0}^{k-1}A_{ik}\left(\sqrt{2} a\right)^{2}
\ee
where we recall the definition of the overlap matrix
\be
A_{ik}\left(v\right) = \int_{-\infty}^{v}\varphi_{i}\left(x\right)\varphi_{k}\left(x\right)dx
\ee
and we used the orthogonality $\int_{-\infty}^{\infty}\varphi_{i}\left(x\right)\varphi_{k}\left(x\right)dx=0$, $i\ne k$.
These equations together lead to Eq.~\eqref{eq:VarNRStep} in the main text with the scaling function
\be
\label{fkEdgeVar}
f_{k}^{\text{edge},\text{Var}}\left(s\right)=\int_{s}^{+\infty}dv\,\left\{ -2\sum_{i=0}^{k-1}A_{ik}\left(v\right)^{2}+A_k\left(v\right)\left[1-A_k\left(v\right)\right]\right\}.
\ee
In particular, $f_{k}^{\text{edge},\text{Var}}\left(s\to +\infty\right)=0$ [because $A_k\left(v\to\infty\right)=1$ and $A_{ik}\left(v\to\infty\right)=0$], matching smoothly with \eqref{eq:VarNR}. Note also that for $f_{k}^{\text{edge},\text{Var}}\left(s\to -\infty\right)=C_{k+1}-C_k$ as can be checked
by comparing with the definition of $C_k$ in \eqref{defC}.

\section{Simplifying $B_{\tilde{R},\Omega}$ and taking the limit $1 - \Omega \ll 1$}
\label{sec:simplifyingB}

We begin by rewriting Eq.~\eqref{eq:BROmega} by expanding the logarithmic term and then performing the integral over the $\lambda$-independent terms:
\bea
B_{\tilde{R},\Omega}&=&\int_{\lambda_{-}(\tilde{R})}^{\lambda_{+}(\tilde{R})}\frac{d\lambda}{2\pi^{2}}\left[\log\left(4\tilde{R}\frac{(1-(\tilde{R}^{2}+\frac{\lambda^{2}}{4\tilde{R}^{2}}-\Omega\lambda))_{+}^{\frac{3}{2}}}{((1+\Omega\lambda)^{2}-\lambda^{2})_{+}^{\frac{1}{2}}}\right)+c_{2}\right] =\frac{1}{2\pi^{2}}\left(\lambda_{+}-\lambda_{-}\right)\left[c_{2}+\log\left(4\tilde{R}\right)\right]\nn\\
&+& \frac{3}{2}\frac{1}{2\pi^{2}}\int_{\lambda_{-}}^{\lambda_{+}}d\lambda\log\left(1-\left(\tilde{R}^{2}+\frac{\lambda^{2}}{4\tilde{R}^{2}}-\Omega\lambda\right)\right)-\frac{1}{2}\frac{1}{2\pi^{2}}\int_{\lambda_{-}}^{\lambda_{+}}d\lambda\log\left((1+\Omega\lambda)^{2}-\lambda^{2}\right) \, .
\eea
The two remaining integrals can in fact be calculated explicitly. The first one gives a simple result:
\bea
\int_{\lambda_{-}}^{\lambda_{+}}d\lambda\log\left(1-\left(\tilde{R}^{2}+\frac{\lambda^{2}}{4\tilde{R}^{2}}-\Omega\lambda\right)\right)&=&\int_{\lambda_{-}}^{\lambda_{+}}d\lambda\log\left(\frac{1}{4\tilde{R}^{2}}\left(\lambda_{+}-\lambda\right)\left(\lambda-\lambda_{-}\right)\right) \nn\\
&=&2\left(\lambda_{+}-\lambda_{-}\right)\left[\log\left(\lambda_{+}-\lambda_{-}\right)-1-\log\left(2\tilde{R}\right)\right] \, ,
\eea
but the second one gives a result that is rather cumbersome:
\be
\label{eq:BIntegralExplicit}
\int_{\lambda_{-}}^{\lambda_{+}}d\lambda\log\left((1+\Omega\lambda)^{2}-\lambda^{2}\right)=\left[\lambda\left(\log\left((\lambda\Omega+1)^{2}-\lambda^{2}\right)-2\right)+\frac{\log(\lambda(\Omega-1)+1)}{\Omega-1}+\frac{\log(\lambda\Omega+\lambda+1)}{\Omega+1}\right]_{\lambda_{-}}^{\lambda_{+}} \, .
\ee
So, using $A_{\tilde{R},\Omega}=\frac{1}{2\pi^{2}}\left(\lambda_{+}-\lambda_{-}\right)$ we reach
\be
\label{eq:BIntegral2}
B_{\tilde{R},\Omega}=A_{\tilde{R},\Omega}\left[c_{2}+\log2+3\log\left(\lambda_{+}-\lambda_{-}\right)-3-2\log\left(2\tilde{R}\right)\right]-\frac{1}{2}\frac{1}{2\pi^{2}}\int_{\lambda_{-}}^{\lambda_{+}}d\lambda\log\left((1+\Omega\lambda)^{2}-\lambda^{2}\right) \; .
\ee

Now we consider the limit $1-\Omega \ll 1$ with $\tilde{R}\sim\left(1-\Omega\right)^{-1/2}\gg1$. 
In this limit, one has $\lambda_{+}\simeq\lambda_{-}\simeq2\tilde{R}^{2}\gg1$, and it is convenient to calculate the integral \eqref{eq:BIntegralExplicit} approximately.
Neglecting terms $O(1)$ or smaller, the term that is inside the logarithm in the integrand in \eqref{eq:BIntegralExplicit} is (using $\lambda\simeq2\tilde{R}^{2}\gg1$)
\be
(1+\Omega\lambda)^{2}-\lambda^{2}\simeq2\lambda\left[1-\left(1-\Omega\right)\lambda\right]\simeq4\tilde{R}^{2}\left[1-2\left(1-\Omega\right)\tilde{R}^{2}\right].
\ee
Under this approximation the integration is trivial because the integrand is a constant:
\be
\int_{\lambda_{-}}^{\lambda_{+}}d\lambda\log\left((1+\Omega\lambda)^{2}-\lambda^{2}\right)\simeq\left(\lambda_{+}-\lambda_{-}\right)\left[2\log\left(2\tilde{R}\right)+\log\left(1-2\left(1-\Omega\right)\tilde{R}^{2}\right)\right].
\ee
Plugging this into \eqref{eq:BIntegral2} yields
\bea
B_{\tilde{R},\Omega} &\simeq& A_{\tilde{R},\Omega}\left[3\log\left(\frac{\lambda_{+}-\lambda_{-}}{2\tilde{R}}\right)-\frac{1}{2}\log\left(1-2\left(1-\Omega\right)\tilde{R}^{2}\right)+c_{2}+\log2-3\right] \nn\\
&\simeq&A_{\tilde{R},\Omega}\left[\log\left(1-2(1-\Omega)\tilde{R}^{2}\right)+c_{2}+4\log2-3\right]
\eea
Finally, using
\be
\label{eq:Aapprox}
A_{\tilde{R},\Omega}=\frac{2}{\pi^{2}}\tilde{R}\sqrt{1-(1-\Omega^{2})\tilde{R}^{2}}\underbrace{\simeq}_{1-\Omega\ll1}\frac{2}{\pi^{2}}\tilde{R}\sqrt{1-2(1-\Omega)\tilde{R}^{2}}
\ee
we get
\be
\label{eq:Bapprox}
B_{\tilde{R},\Omega}\simeq\frac{2}{\pi^{2}}\tilde{R}\sqrt{1-2(1-\Omega)\tilde{R}^{2}}\left[\log\left(1-2(1-\Omega)\tilde{R}^{2}\right)+c_{2}+4\log2-3\right] \; .
\ee
Plugging Eqs.~\eqref{eq:Aapprox} and \eqref{eq:Bapprox} into \eqref{eq:VarNROmegaO1}, we reach
\be
\label{eq:VarNRMatching2}
{\rm Var}\,{\cal N}_{R}\simeq\frac{2}{\pi^{2}}\tilde{R}\sqrt{1-2(1-\Omega)\tilde{R}^{2}}\mu\left[\log\mu+\log\left(1-2(1-\Omega)\tilde{R}^{2}\right)+c_{2}+4\log2-3\right]
\ee
which agrees with \eqref{eq:VarNRMatching}.

\section{Details of numerical simulations}
\label{app:sim}

Here we briefly describe the method that we used in order to simulate the radial coordinates $r_1 , \dots , r_N$ of the particles for the rotating HO in 2d, which we then used in order to empirically measure the densities and number variances displayed in Figs.~\ref{FigDensity} and \ref{FigVarNR} in the main text.
We use the decoupling between angular sectors, see \cite{UsCounting2020} for details, so that we generate samples of radial coordinates within each angular sector $\ell$ independently. For each sector we generate samples of the positions of $m_\ell$ fermions in the effective 1d potential $V_\ell(r)= \frac{r^2}{2} + \frac{\ell^2-\frac{1}{4}}{2 r^2} - \Omega \ell$ [which is simply \eqref{Vell} for the HO], where $m_\ell$ is given by \eqref{eq:mell}.
This is conveniently done by exploiting the mapping between the effective 1d systems to the eigenvalues of a random matrix from the Wishart-Laguerre Unitary Ensemble  (LUE) -- this mapping is described, e.g., in the recent review \cite{DeanReview2019}. Finally, we used the tridiagonal matrix representations of Gaussian unitary ensemble (GUE) and LUE matrices \citep{Dumitriu2002} in order to efficiently generate their eigenvalues.
The code that we used is avaliable in \cite{SM}.

\section{Higher cumulants}
\label{sec:higher} 

Here we sketch how one would proceed to compute the higher cumulants of ${\cal N}_R$. Ideally we would like to obtain the FCS generating function. By the same arguments as in the main text it will be a piecewise
linear function of $R$. In the interval in $\sqrt{\ell_{k}} < R < \sqrt{\ell_{k-1}}$ associated to the Landau level indexed by $n=k-1$ 
\bea \label{FCS1} 
\log\left\langle e^{-s\left({\cal N}_{R}-\left\langle {\cal N}_{R}\right\rangle \right)}\right\rangle &\simeq& \sqrt{2}R\int_{-\infty}^{\infty}da\log\left\langle e^{-s\left({\cal N}_{\left(-\infty,a\right]}-\left\langle {\cal N}_{\left(-\infty,a\right]}\right\rangle \right)}\right\rangle \nn\\
&=&\sqrt{2}R\int_{-\infty}^{\infty}da\left(\log\det\left[\delta_{ij}-\left(1-e^{-s}\right)A_{ij}(a)\right]+s\,{\rm Tr}A(a)\right)
\eea 
where ${\cal N}_{\left(-\infty,a\right]}$ is the number of fermions in $\left(-\infty,a\right]$ for the standard harmonic oscillator (HO)
$H = \frac{p^2}{2} + \frac{x^2}{2}$,
and the trace and determinant are over the eigenstates $i,j=0,\dots,n$ of the HO.
Note that we had to substract the mean, since $\int da\left\langle {\cal N}_{\left(-\infty,a\right]}\right\rangle =+\infty$
(a manifestation of the fact that all cumulants are piecewise linear in $R$, except the mean which has a quadratic dependence $R^2$). 
In the second line we have used the standard formula for the FCS of a determinantal process. The overlap 
matrix $A_{ij}(a)$ is defined in \eqref{eq:defA}. Computing the integrals and 
expanding \eqref{FCS1} in powers of $s$ 
allows to obtain the $p$-th cumulant as the coefficient of $s^p/p!$. 

Let us give some examples of this method. One has for $n=k-1=0$
\bea
&& \frac{1}{\sqrt{2}R}\log\left\langle e^{-s\left({\cal N}_{R}-\left\langle {\cal N}_{R}\right\rangle \right)}\right\rangle \simeq\int_{-\infty}^{\infty}da\left[\log\left(1-\frac{1-e^{-s}}{2}\left(1+{\rm erf}(a)\right)\right)+\frac{s}{2}\left(1+{\rm erf}(a)\right)\right] \\
&& = 
\frac{s^{2}}{2}\frac{1}{\sqrt{2\pi}}+\frac{s^{4}}{4!}\int_{-\infty}^{\infty}\frac{-3\text{erf}(a)^{4}+4\text{erf}(a)^{2}-1}{8}\,da+\frac{s^{6}}{6!}\int_{-\infty}^{\infty}\frac{-15\text{erf}(a)^{6}+30\text{erf}(a)^{4}-17\text{erf}(a)^{2}+2}{8}\,da+O(s^{8}) \nn \\
\label{eq:cumulants246}
&&=\frac{s^{2}}{2}\frac{1}{\sqrt{2\pi}}+\left[\frac{9\sqrt{2}}{\pi^{3/2}}\text{arctan}\left(\frac{1}{\sqrt{8}}\right)-\sqrt{\frac{2}{\pi}}\right]\frac{s^{4}}{4!}+0.00893625\frac{s^{6}}{6!}+O(s^{8})
\eea
which is in agreement with the results for the first three cumulants of \cite{LMG19}.
We calculated the fourth cumulant as follows. We rewrite the coefficient of $s^{4}/4!$ as  $(4I_{2}-3I_{4})/8$
where we defined $I_{m}=\int_{-\infty}^{\infty}\left(\text{erf}\left(x\right)^{m}-1\right)dx$.
A direct integration using Mathematica gives $I_{2}=-2\sqrt{2/\pi}$. We now calculate $I_4$. We define
\be
\mathcal{I}_4 \left(\alpha_{1},\alpha_{2},\alpha_{3},\alpha_{4}\right) =\int_{-\infty}^{\infty}\left[\text{erf}\left(\alpha_{1}x\right)\text{erf}\left(\alpha_{2}x\right)\text{erf}\left(\alpha_{3}x\right)\text{erf}\left(\alpha_{4}x\right)-1\right]dx \, .
\ee
In order to convert the integral from an integral over error functions to a Gaussian integral, we take partial derivatives with respect to the $\alpha_i$'s:
\be
\frac{\partial^{4}\mathcal{I}_{4}}{\partial\alpha_{1}\partial\alpha_{2}\partial\alpha_{3}\partial\alpha_{4}}=\left(\frac{2}{\sqrt{\pi}}\right)^{4}\int_{-\infty}^{\infty}x^{4}e^{-\left(\alpha_{1}^{2}+\alpha_{2}^{2}+\alpha_{3}^{2}+\alpha_{4}^{2}\right)x^{2}}dx=\left(\frac{2}{\sqrt{\pi}}\right)^{4}\frac{3\sqrt{\pi}}{4\left(\alpha_{1}^{2}+\alpha_{2}^{2}+\alpha_{3}^{2}+\alpha_{4}^{2}\right)^{5/2}} \, .
\ee
This equation can now be integrated using Mathematica, and it yields
\be
\label{calI4sol}
\mathcal{I}_{4}\left(\alpha_{1},\alpha_{2},\alpha_{3},\alpha_{4}\right)=-\frac{1}{3}\left(\frac{2}{\sqrt{\pi}}\right)^{4}\frac{3\sqrt{\pi}}{4}\sum_{\left\{ i,j\right\} \subset\left\{ 1,2,3,4\right\} }\frac{\sqrt{\alpha_{i}^{2}+\alpha_{j}^{2}}}{\alpha_{i}\alpha_{j}}\text{arctan}\left(\frac{\alpha_{1}\alpha_{2}\alpha_{3}\alpha_{4}}{\alpha_{i}\alpha_{j}\sqrt{\alpha_{i}^{2}+\alpha_{j}^{2}}\sqrt{\alpha_{1}^{2}+\alpha_{2}^{2}+\alpha_{3}^{2}+\alpha_{4}^{2}}}\right)\,,
\ee
where the sum is over all $\left(\begin{array}{c}
4\\
2
\end{array}\right)$ ways to choose a subset $\left\{ i,j\right\} $ of the indices $\left\{ 1,2,3,4\right\} $ \cite{footnote:I4}.
%
 Finally, Eq.~\eqref{calI4sol} gives $I_{4}=\mathcal{I}_{4}\left(1,1,1,1\right)= -\frac{24\sqrt{2}}{\pi^{3/2}}\text{arctan}\left(\frac{1}{\sqrt{8}}\right)$.
The values of $I_2$ and $I_4$ yield the coefficient of $s^{4}/4!$ in \eqref{eq:cumulants246}.

One has for $n=k-1=1$ 
\bea
&&  \frac{1}{\sqrt{2}R}\log \left\langle e^{-s\left({\cal N}_{R}-\left\langle {\cal N}_{R}\right\rangle \right)}\right\rangle  \nn\\
&& \simeq  \int_{-\infty}^{\infty} da \left[
\log \det \left(
\begin{array}{cc}
 1-\frac{1}{2} \left(1-e^{-s}\right) (\text{erf}(a)+1) & \frac{e^{-a^2}
   \left(1-e^{-s}\right)}{\sqrt{2 \pi }} \\
 \frac{e^{-a^2} \left(1-e^{-s}\right)}{\sqrt{2 \pi }} & 1-\frac{1}{2}
   \left(1-e^{-s}\right) \left(-\frac{2 e^{-a^2} a}{\sqrt{\pi }}+\text{erf}(a)+1\right) 
   \\
\end{array}
\right) + s \, {\rm Tr} A(a) \right] \nn\\
&& =  \int_{-\infty}^{\infty} da \bigg[
\log
\frac{e^{-2 \left(a^2+s\right)} \left(2 \sqrt{\pi } a e^{a^2} \left(e^s-1\right)
   \left(\text{erf}(a)+e^s \text{erfc}(a)+1\right)+\pi  e^{2 a^2} \left(\text{erf}(a)+e^s
   \text{erfc}(a)+1\right)^2-2 \left(e^s-1\right)^2\right)}{4 \pi } \nn \\
   && + 
   s \left(-\frac{e^{-a^2} a}{\sqrt{\pi }}+\text{erf}(a)+1\right) \bigg] \nn\\
&& = \frac{s^2}{2} \frac{7}{4 \sqrt{2 \pi }} + \frac{s^4}{4!} 
 \int_{-\infty}^{\infty} \bigg[
\frac{a e^{-a^2} \left(3 \text{erf}(a)^2-2\right) \text{erf}(a)}{\sqrt{\pi }}+\frac{6 a
   \left(2 a^2+3\right) e^{-3 a^2} \text{erf}(a)}{\pi ^{3/2}}-\frac{\left(a^2+1\right)
   e^{-2 a^2} \left(9 \text{erf}(a)^2-2\right)}{\pi } \nn\\
   && -\frac{3 \left(2 a^4+4 a^2+1\right)
   e^{-4 a^2}}{\pi ^2}-\frac{3}{4} \text{erf}(a)^4+\text{erf}(a)^2-\frac{1}{4} \bigg] + O(s^6) \nn\\
 && = \frac{s^2}{2} \frac{7}{4 \sqrt{2 \pi }} -0.0322399 \frac{s^4}{4!} + O(s^6)
\eea
in agreement with our result for $C_2$ and giving the fourth cumulant for $n=k-1=1$. 
We have also checked that the method using the generating functions and Mehler's formula can be used,
but it leads to cumbersome integrals. 

\end{widetext}

{}


\begin{thebibliography}{}
\bibitem{footnote:pCumulant}
In Eq.~\eqref{cumulants0} and in other places in the paper, the superscript $p$ denotes the index of the cumulant, and is not to be confused with the momentum $p$.
\bibitem{footnote:rescaling}
Note that for a 1d Hamiltonian $H_{\hbar,\omega}= \frac{\hbar^2}{2} \partial_x^2 + \frac{1}{2} \omega^2 x^2$ one has for $m$ fermions
$\left.{\cal N}_{[a,b]}\,\right|_{H_{\hbar,\omega},m}=\left.{\cal N}_{[\alpha a,\alpha b]}\,\right|_{H_{1,1},m}$, $\alpha=\sqrt{\hbar/\omega}$
hence here $\alpha=\sqrt{2}$ which leads to the relation \eqref{eq:Rescaling} in the text.
\bibitem{footnote:I4}
We used the asymptotic behavior of $\mathcal{I}_{4}\left(\alpha_{1},\alpha_{2},\alpha_{3},\alpha_{4}\right)$ at $\alpha_i \to0$  to determine that no sums of functions of three out of the four $\alpha_i$'s should be added to the right hand side of \eqref{calI4sol}.
For instance, the lack of a term of order $\alpha_{4}^{0}$ in the asymptotic expansion of $\mathcal{I}_{4}\left(\alpha_{1},\alpha_{2},\alpha_{3},\alpha_{4} \to 0\right)$ rules out the addition of an arbitrary function of $\alpha_{1},\alpha_{2},\alpha_{3}$ in \eqref{calI4sol}.
\end{thebibliography}
\end{document}